\documentclass[aps,twocolumn,showpacs]{revtex4}
\usepackage{amsmath}
\usepackage{graphicx}

\newcommand{\p}[1]{\phantom{#1}}
\newcommand{\be}{\begin{equation}}
\newcommand{\ee}{\end{equation}}
\newcommand{\bea}{\begin{eqnarray}}
\newcommand{\eea}{\end{eqnarray}}

\newcommand{\bi}{\begin{itemize}}
\newcommand{\ei}{\end{itemize}}
\newcommand{\grad}{\boldsymbol{\nabla}}

\newcommand{\teves}{TeVeS}

\begin{document}

\def\araa{{\em ARAA}}
\def\aj{{\em AJ}}
\def\apj{{\em ApJ}}
\def\apjl{{\em ApJ}}
\def\apjs{{\em ApJS}}
\def\aap{{\em A\&A}}
\def\mnras{{\em MNRAS}}

\title{TeVeS gets caught on caustics}
\author{Carlo R. Contaldi}\email{c.contaldi@imperial.ac.uk}%
\author{Toby Wiseman}\email{t.wiseman@imperial.ac.uk}
 \author{Benjamin Withers}
\email{benjamin.withers02@imperial.ac.uk}
 \affiliation{%
Theoretical Physics, Blackett Laboratory, Imperial College, Prince
 Consort Road, London, SW7 2BZ, U.K.}%

\date{\today}
\begin{abstract}
  TeVeS uses a dynamical vector field with timelike unit norm
  constraint to specify a preferred local frame. When matter moves
  slowly in this frame - the so-called quasi-static regime - Modified
  Newtonian Dynamics (MoND) results.  Theories with such vectors (such
  as Einstein-Aether) are prone to the vector dynamics forming
  singularities which render their classical evolution
  problematic. Here we analyse the dynamics of the vector in TeVeS in
  various situations.  We begin by analytically showing that the
  vacuum solution of TeVeS forms caustic singularities under a large
  class of physically reasonably initial perturbations. This shows the
  classical evolution of TeVeS appears problematic in the absence of
  matter.  We then consider matter by investigating black hole
  solutions. We find large classes of new black hole solutions with
  static geometries where the curves generated by the vector field are
  attracted to the black hole and may form caustics. We go on to
  consider the full dynamics with matter by numerically simulating,
  assuming spherical symmetry, the gravitational collapse of a scalar,
  and the evolution of an initially nearly static boson star. We find
  that in both cases our initial data evolves so that the vector field
  develops caustic singularities on a time scale of order the
  gravitational in-fall time.  Having shown singularity formation is
  generic with or without matter, Bekenstein's original formulation of
  TeVeS appears dynamically problematic.   We argue that by modifying the
  vector field kinetic terms to the more general form used by
  Einstein-Aether this problem may be avoided.
\end{abstract}


\maketitle

\section{Introduction}
For many years the existence of dark matter has been postulated to
reconcile a number of astrophysical and cosmological observations with
our understanding of the laws of gravitation. Dark matter was
originally introduced decades ago to explain the discrepancy between
the rotation velocities of stars in the outskirts of galaxies and that
predicted by the mass inferred from the amount of visible mass in the
galaxy.

The success of the dark matter paradigm extends beyond galactic scales
to cluster and indeed cosmological scales. Today we know that best fit
models of structure formation apparently require a dark matter
fraction much larger than the known baryon content of the universe to
drive the growth of structure from kpc through to Gpc scales. The
potential wells provided by a cold, dark matter (CDM) component
also reconcile the amplitude of the acoustic peaks observed in the
Cosmic Microwave Background (CMB) angular power spectrum \cite{boom,dasi,cbi,wmap}
with the known baryon content.

Dark matter also provides a simple explanation for the observed lensing
of background galaxies by clusters along the line of sight. Recently
the combination of optical, x-ray and lensing observations of the
bullet cluster have yielded the most direct evidence to date in
support of the picture where the gravitational mass of clusters is
dominated by a dark matter component \cite{Clowe06}. 

Taken as a whole the growing wealth of observations points clearly to
a concordance $\Lambda$CDM cosmological model with a significant
fraction of the critical energy density made up of CDM. The dark
matter paradigm has stood the test of time remarkably well but
significant questions remain. Many candidates for a dark matter
particle exist ranging from massive neutrinos to more exotic weakly
interacting extensions to the standard model. However dark matter has
yet to be detected directly in the laboratory or indirectly possibly
through the $\gamma$-ray signature of its decay in the centre of
galaxies (this is required to avoid the concentration of dark matter
observed in numerical simulations).

For these reasons an alternative approach to adding a dark matter
component has been to consider whether the discrepancies between
observations and general relativity in the low acceleration regime are
an indication of the failure of the theory itself. This was the
approach taken by Milgrom \cite{Milgrom83} who proposed a
phenomenological modification to the acceleration equation which seems
to fit well galactic rotation curves without the addition of any dark
matter:
\begin{equation}\label{eq:mond}
  \mu\left( |{\bf a}|/a_0\right){\bf a} = -\nabla \Phi,
\end{equation}
where $\Phi$ is the Newtonian potential, $\mu(x)$ is an arbitrary
function with limits such that $\mu(x)\rightarrow 1$ in the strong
acceleration regime ($x \gg 1$). The constant $a_0 \approx 10^{-10}$
m~s$^{-2}$ determines the acceleration scale below which the Modified
Newtonian Dynamics or MoND becomes relevant, and the above
acceleration law receives non-linear corrections. While $\mu$ is
potentially a free (monotonic) function, only the limits
where its argument goes to zero or infinity affect the
astrophysical phenomenology.

MoND has been successful in fitting the anomalous accelerations
observed in galaxies and clusters (see e.g. \cite{Sanders02} for a
recent review). It also successfully predicts the Tully-Fisher relation
correlating the luminosity of galaxies to the fourth power of the
rotation velocity. However it remains a phenomenological modification
of gravity with no underlying relativistic theory. In addition the
simplest theory based on a modified action that reproduces
Eq~\ref{eq:mond} depends explicitly on coordinates and breaks
conservation laws. The fact that MoND has no underlying covariant
theory has restricted its application for the purpose of comparison
with other astrophysical and cosmological observations. For example,
lensing predictions in both the strong and weak regimes cannot be
formulated in MoND and this has left it unable to answer the criticism
stemming from lensing mass reconstructions of galactic and cluster
profiles which seem to suggest the existence of dark matter halos.

Recently however, Bekenstein \cite{Bekenstein:2004ne}, has put forward
\teves, a relativistic theory of gravity which reduces to MoND in the
weak acceleration limit. In \teves\ the matter sector lives on a
matter frame (MF) metric which maps `disformally' to a second,
Einstein or gravitational, frame (EF) metric via a dynamical scalar
field $\phi$ and a dynamical vector field $A$. The addition of a
scalar and vector degree of freedom are behind the name `Tensor,
Vector, and Scalar', or \teves\ theory. \teves\ builds on previous
attempts to obtain a relativistic version of MoND which suffered from
a number of inconsistencies involving the acausal propagation of
physical degrees of freedom \cite{Bekenstein94,Sanders97}. \teves\
however was shown to be a fully causal theory for positive values of
the additional scalar field. 

The original motivation behind \teves\ was to build a theory with a
fully consistent action which recovers the MoND behaviour in the weak
acceleration limit. However, given it is a relativistic, metric theory
of gravity and matter, it can do much more. In \teves\ it is possible
to calculate geodesics in the presence of a matter sources which leads
to lensing predictions \cite{Zhao06}. It is also possible to
show that it is compatible with the basic background cosmological
observations such as age and distance measure observations
\cite{Bekenstein:2004ne}. The full framework of relativistic
perturbation theory can be developed in \teves\ which makes comparison
to the perturbed universe possible. Already the first calculations in
this area have shown that the theory may be reconciled with CMB and
Large Scale Structure (LSS) observations
\cite{Skordis:2005eu,Dodelson06,Bourliot07,Skordis:2008pq}, albeit
with some fine tuning of the model ingredients. Attempts have also
been made to explain the bullet cluster results within the \teves\
framework \cite{Angus07}.

For \teves\ to be a successful theory it must also be shown to be
consistent, and agree with observations, in the strong gravity
regime. In exploring this end of the theory the potential is that it
could be compared to astrophysical observations of compact objects
such as neutron stars and black holes or at the solar system level
with post Newtonian (ppn) corrections to planetary orbits
\cite{Bekenstein:2004ne,Giannios:2005es,Bekenstein:2006fi}.

In order to have a modification of gravity dependent on acceleration,
one must have a reference frame in which to measure that acceleration.
The vector in TeVeS dynamically selects that reference frame,
spontaneously breaking Lorentz invariance since it is constrained to
have unit timelike norm. All types of matter see the same distorted
metric so adding a preferred frame is not in conflict with weak
equivalence principle tests. Only tests of gravitational dynamics can
constrain the theory. MoND is recovered from TeVeS when matter moves
non-relativistically in the frame defined by the TeVeS vector, which
has been termed the `quasi-static' regime.  The purpose of our work is
to argue that this quasi-static regime will typically only exist for a
short period of time, of order the gravitational in-fall time, after
which the vector field develops a singularity and the theory cannot be
classically evolved any further. Hence TeVeS even classically is
dynamically sick in practice and recovery of MoND or even GR is
impossible. Indeed here for simplicity we will focus on the large
acceleration regime relevant on small scales (e.g. within the solar
system) where the in-fall time scales are shortest.

This singular vector field behaviour is analogous to that in other
modified gravity theories such as Einstein-\AE ther theory
\cite{Jacobson:2000xp,Clayton:2001vy} and Ghost Condensation
\cite{ArkaniHamed:2003uy,ArkaniHamed:2005gu}. Einstein-\AE ther theory
is much simpler than TeVeS, being simply Einstein gravity modified by
adding a vector field, again with timelike unit norm constraint. The
vector action is taken to be more general than that in TeVeS where it
is simply that of a Maxwell field, but one may choose them to be the
same. In this case (actually a theory written down earlier
\cite{Kostelecky:1989jw}), it is easy to show that the vector field
generically develops singularities; classes of solutions exist where
the integral curves of the vector are timelike geodesics moving in the
spacetime geometry created by the matter. These geodesics fall into
gravity potential wells and meet, and when they do so, the flow they
define develops caustic singularities \cite{Jacobson:2000xp}. The
vector field at these points becomes singular. It is for this reason
that the Einstein-\AE ther literature focuses on other choices of the
vector action than Maxwell type. Indeed while the ghost condensation
theory has no vector, it is the integral curves of the gradient of the
ghost scalar that form caustics. Since TeVeS is a considerably more
complicated theory than Einstein-\AE ther, with complicated coupling
of its vector and scalar to the matter, the vector behaviour and in
particular whether it forms singularities could be very different. Our
key result is that while in detail the dynamics of the vector is
clearly different, it is still subject to the same singularity
development as the Maxwell case of Einstein-\AE ther. However, we can
play the same game as in Einstein-\AE ther theory, and by taking more
general vector kinetic terms, we may avoid this behaviour, and as we
show later, we still recover MoND for quasi-static systems.

The {\sl paper} is organised as follows. In section~\ref{action} we
review \teves\ theory and the field equations derived from the \teves\
action. We review the relation between TeVeS and the Einstein-\AE ther
theory in section~\ref{aether} where we introduce the problem of
vector field caustic singularity formation. In
section~\ref{analyticshock} we begin by showing analytically that in
the absence of matter the TeVeS vacuum forms singularities under
evolution of a class of initial perturbations. These are physically
reasonable perturbations and we are able to precisely characterize the
condition for the initial data developing to a singularity. We find
that the condition is generic within the class of
perturbations. Already this analysis indicates a fundamental problem
with the TeVeS dynamics. However, phenomenologically it is the
inclusion of matter in TeVeS that is of key interest, and hence we
proceed to study whether such inclusion ameliorates or worsens the
problem. We begin our study of matter in TeVeS in section~\ref{bh}
with an analytic treatment of black holes.  Previous static black hole
solutions of Giannios \cite{Giannios:2005es} and Sagi and Bekenstein
\cite{Sagi:2007hb} have a static vector field aligned with Killing
time. However we find a large class of new solutions where the
geometry is static but the vector field is in general dynamic. There
exists a family of stationary solutions where the vector falls into
the future horizon, but also dynamic solutions where caustics may form
in the exterior of the black hole. The black hole provides a focus for
the curves of the vector field and hence indicates that singularity
formation is likely to be enhanced by compact matter sources. We then
proceed in section~\ref{shockformation} with our study of matter by
performing full numerical simulation of scalar collapse and the
evolution of an initially quasi-static boson star. In both cases the
evolution ends at a caustic, in the former outside of an apparent
horizon, and in the latter near the surface of the star.  Having given
evidence that the dynamics of \teves\ is too pathological
to provide a relativistic setting for the phenomenological theory of
MoND, in section~\ref{modsection} we outline a modified \teves\ theory
which may not suffer from caustic formation in the vector field. We
explicitly show that MoND is again reproduced in the appropriate
Newtonian limit. We argue that this modification is likely to affect
many detailed phenomenological studies of \teves\ should therefore be
included.  We conclude with a summary and brief discussion of our main
results in section~\ref{disc}.


\subsection{TeVeS Action and Field Equations}\label{action}
TeVeS is constructed using two metrics, the matter frame (MF) metric
$\tilde{g}$, and the Einstein frame (EF) metric $g$. The two metrics
are related through `disformal relations' involving the extra scalar and
vector fields $\phi$ and $A_\mu$
\begin{eqnarray}\label{disformal}
\tilde{g}_{\alpha \beta} &=& e^{-2\phi} g_{\alpha \beta} - 2 A_\alpha
A_\beta \sinh 2 \phi, \\ 
\tilde{g}^{\alpha \beta} &=& e^{2\phi}
g^{\alpha \beta} + 2 A^\alpha A^\beta \sinh 2 \phi.
\end{eqnarray}
The total action $S$ governing the dynamics in \teves\ can be split
into separate components $S=S_{g+v} + S_s +S_m$, where
\begin{eqnarray}\label{tevesvectoraction} 
S_{g+v} &=&\frac{1}{16\pi G}\int
\left[R-\frac{K}{2}F_{\mu\nu}F^{\mu\nu} +\right.\nonumber\\ 
&&\biggr. \lambda(A^2 +1)\biggr]  \bigr) (-g)^{1/2} d^4 x, 
\end{eqnarray}
where $g$ is the determinant of the EF metric, $R$ is the scalar
curvature, $G$ is the gravitational constant and
$F_{\mu\nu}=A_{[\mu,\nu]} = A_{\mu,\nu} - A_{\nu,\mu}$. The Lagrange multiplier
$\lambda$ enforces the timelike, unit norm constraint on the vector
field
\begin{equation}
g^{\mu\nu}A_{\mu}A_{\nu} = -1.
\end{equation}
The scalar field action is given by
\begin{eqnarray} 
S_s &=&-\frac{1}{2}\int\biggl[\sigma^2
(g^{\alpha\beta}-A^\alpha
A^\beta)\phi_{,\alpha}\phi_{,\beta}+\biggr.\nonumber\\
&&\left.\frac{1}{2}G \ell^{-2}\sigma^4 {\cal F}(\kappa G\sigma^2) \right](-g)^{1/2} d^4 x,
\label{tevesscalaraction}
\end{eqnarray}
where $\sigma$ is a non-dynamical scalar field and ${\cal F}(\kappa
\,G\, \sigma^2)$ is a dimensionless function whose behaviour is
determined by requiring GR and MoND to be recovered in the appropriate
dynamical limits. \teves\ introduces three new parameters; the two
dimensionless constants $\kappa$ and $K$ and a third parameter $l$
with units of length.

Finally the matter action $S_m$ is built using the MF metric as
\begin{equation}
  S_m = \int L\left[ \tilde{g}, \chi^A, \partial \chi^A\right](-\tilde{g})^{1/2} d^4 x,
\end{equation}
for a collection of matter fields $\chi^A$. Thus all matter fields are
coupled to the same MF metric and test particles follow the same
geodesics. This ensures the weak equivalence principle is satisfied
and the theory is not in conflict with fifth force measurements. 

Varying with respect to $g$, and
recalling $\tilde{g}=\tilde{g}\,(g,A,\phi)$ gives the Einstein equations 
\begin{equation}
 G_{\alpha \beta}=8\pi G[\tilde{T}_{\alpha
\beta}+(1-e^{-4\phi})A^{\mu} \tilde{T}_{\mu(\alpha}
A_{\beta)}+\tau_{\alpha \beta}]+\Theta_{\alpha \beta},
\label{metrice}
\end{equation}
where $\tilde{T}_{\mu(\alpha}A_{,\beta)} = \tilde{T}_{\mu\alpha} A_{\beta} +\tilde{T}_{\mu\beta} A_{\alpha}$, $G_{\alpha \beta}$ is the Einstein tensor, $\tilde{T}_{\alpha
\beta}$ is the energy momentum tensor of the matter components defined
in terms of the MF metric $\tilde{g}_{\alpha\beta}$ and
\begin{widetext}
\begin{equation}
\tau_{\alpha\beta} \equiv
\sigma^2\left[\phi_{,\alpha}\phi_{,\beta}-{ 1\over
 2}g^{\mu\nu}\phi_{,\mu}\phi_{,\nu}\,g_{\alpha\beta}-
A^\mu\phi_{,\mu}\left(A_{(\alpha}\phi_{,\beta)}- { 1\over
 2}A^\nu\phi_{,\nu}\,g_{\alpha\beta}\right)\right]
-{ 1\over 4}G \ell^{-2}\sigma^4
{\cal F}(\kappa G\sigma^2)  g_{\alpha\beta},
\end{equation}\label{tau}
\end{widetext}
\begin{equation} 
\Theta_{\alpha\beta}\equiv K\left(F_{\alpha}^{\p{\alpha}\mu}
F_{\beta \mu}-{ 1\over  4} g_{\alpha\beta} F^2
\right)- \lambda A_\alpha A_\beta.
\label{Theta}
\end{equation}
Variation with respect to the scalar field $\sigma$ yields a relation between
$\sigma$ and $\phi_{,\alpha}$ involving ${\cal F}$. The specific choice
of ${\cal F}$ determines the exact behaviour of the theory in the weak
acceleration regime and is relevant for the MoND and cosmological
behaviour of \teves.  The regime of interest for this work is one
where the acceleration is much stronger than the MoND acceleration
scale $a_0$. In this case the MoND function $\mu(|a|/a_0)\rightarrow
1$, which is equivalent to a limit on the argument of the free function
${\cal F}$, 
\begin{equation}
\sigma^2\rightarrow\frac{1}{\kappa G}.
\end{equation}
For any suitable function choice, ${\cal F}$ diverges logarithmically
in $(\mu-1)$ in this limit. The contribution of $\mathcal{F}$ to the field equations
(\ref{metrice}) however is suppressed by a factor $(\mu-1)$ relative to other terms, and so when $\mu\sim 1$ it may be neglected \cite{Bekenstein:2004ne,Giannios:2005es,Sagi:2007hb}. Thus
our results will be insensitive to any particular choice of ${\cal F}$
and we drop the term in the following. Finally variation with respect
to the scalar gives
\begin{eqnarray} 
\left[(g^{\alpha \beta} - A^\alpha A^\beta) \phi_{,\alpha}
\right]_{;\beta} =&&  \nonumber\\
\kappa G \left[g^{\alpha \beta} +(1+ e^{-4
\phi})A^\alpha A^\beta \right]\tilde{T}_{\alpha \beta},&& 
\end{eqnarray}
 and for the vector we have
\begin{eqnarray} K \nabla_\beta F^{\beta \alpha} + \lambda A^{\alpha} +
\frac{8\pi}{\kappa} A^{\beta} \phi_{,\beta} g^{\alpha \gamma}
\phi_{,\gamma} = \nonumber\\
8\pi G (1-e^{-4 \phi}) g^{\alpha \mu} A^{\beta}
\tilde{T}_{\mu \beta}.
\end{eqnarray}

As stated here \teves\ is a classical phenomenological theory. The
somewhat Baroque form for the Lagrangian leads to the obvious concern
that the theory is not stable to quantum corrections. Attempts have
been made to study a UV origin from String
theory~\cite{Mavromatos:2007xe,Mavromatos:2007sp}, and there are
certainly many interesting questions in these directions which we do
not consider here.

It is also worth mentioning that the \teves\ theory itself been
generalized by various authors
\cite{Bonvin:2007ap,Zlosnik:2006zu,Sanders:2005vd,Bruneton:2006gf,Skordis:2008pq}
and it would be interesting to consider the formation of caustics
which we study here in these modified versions of the theory.

\subsection{ \AE ther theory, its relation to TeVeS,  and problems with its vector field dynamics}\label{aether}

Another theory of aether field dynamics is Einstein-\AE ther theory
\cite{Jacobson:2000xp} - an effective field theory designed to
investigate the effects of Lorentz violation in a fully covariant
setting. It has the action,
\begin{eqnarray} 
\frac{1}{16\pi G}\int \left[R + K^{\alpha \beta}_{\p{\alpha \beta} \mu \nu}
\nabla_\alpha A^\mu \nabla_\beta A^\nu +\right.&&\nonumber\\
\Bigl.\lambda \left( A^2 +1 \right)\Bigr](-{g})^{1/2} + \int \mathcal{L}_{matter}[g] ,&&
\end{eqnarray}
where $K^{\alpha \beta}_{\p{\alpha \beta} \mu \nu}$ provides the most
general kinetic term for $A$ which is diffeomorphism invariant,
quadratic in derivatives and (preemptively) consistent with the $A^2
=-1$ constraint. Specifically,
\begin{equation} K^{\alpha \beta}_{\p{\alpha \beta} \mu \nu} = c_1
  g^{\alpha\beta} g_{\mu\nu} + c_2 \delta^\alpha_\mu \delta^\beta_\nu
  + c_3 \delta^\alpha_\nu \delta^\beta_\mu + c_4 A^\alpha A^\beta
  g_{\mu\nu} . 
\end{equation} This kinetic term is the usual Maxwell
case when $c_+\equiv c_1+c_3=0, c4=c2=0, c_-\equiv c_1-c_3 <
0$. Einstein-\AE ther theory is actually a truncation of TeVeS in the
absence of matter, where we may consistently set the scalar to zero
and then $c_-=-2K$ - however obviously phenomenologically this is not
the regime of interest for TeVeS where the coupling to matter and the
non-zero scalar are crucial.

Following Jacobson \& Mattingly \cite{Jacobson:2000xp} it is easy to
see that the Maxwell case of Einstein-\AE ther is pathological. To any
solution of Einstein gravity coupled to matter, we may simply add a
vector field obeying the equations,
\begin{equation}\label{AEsoln}
F_{\mu\nu} = 0, \qquad A^2 = -1,
\end{equation}
and this will then solve the full Einstein-\AE ther equations for that
matter since the vector and constraint contribute nothing to the
stress energy. Note that the vacuum, the Minkowski geometry and $A^\mu
= (\partial_t)^\mu$, is in this class of solutions. Generally the
solution is given by,
\begin{equation}\label{AEsoln2}
A_\mu = \partial_\mu \chi , \qquad (\partial \chi)^2 = -1,
\end{equation}
where the latter equation is a p.d.e., first order in time, 
\begin{equation}
\partial_t \chi = \frac{1}{(- g^{tt})} \left( g^{ti} \partial_i \chi - \sqrt{ (- g^{tt}) ( 1 + \partial^i \chi \partial_i \chi) + ( g^{ti} \partial_i \chi)^2} \right),
\label{eq:gaugepde}
\end{equation}
with $i = 1,\ldots,3$, which can evolve $\chi$ in time from an initial
Cauchy surface. We have taken the choice of root above since we wish
$A^\mu$ to be a future directed vector field. Hence the data for the
solutions can be characterized by the function $\chi(t=0,x)$.  Now,
$A^\nu \nabla_\nu A_\mu = A^\nu \nabla_\mu A_\nu = 0$ using both
relations in \eqref{AEsoln}. Hence integral curves of the vector field
$A$ are simply timelike geodesics.

\begin{figure}[t]
\begin{center}
\includegraphics[angle=0,width=2.9in]{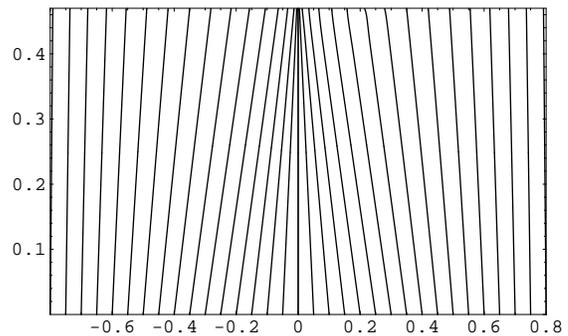}
\caption{\emph{Flat Space} evolution of a Gaussian perturbation to the
  radial component of the vector field for the Einstein-Maxwell case
  ($c_2=0$, no divergence term included). The solution displays
  caustic instabilities.}
\label{flatplot1}
\end{center}
\end{figure}

Suppose we consider a static star as a matter source. Then the
solution above will have families of vector fields with different
initial directions, but all will have integral curves that fall in
towards the gravitational potential well and will meet each other in a
timescale of order the gravitational in-fall time. Since these are
integral curves of the vector field, when they meet they result in a
caustic singularity where the value of the vector is ill-defined.
Indeed, even in the absence of matter and with the Minkowski spacetime
geometry it is possible to have singular behaviour. Such a solution is
illustrated in figure~\ref{flatplot1}.

This simple argument shows for solutions with $F_{\mu\nu} = 0$ that
caustic singularities generically occur in the presence of
gravitational potential wells. It seems reasonable that singularities
will also occur in solutions where $F_{\mu\nu} \ne 0$. Whilst there
are no general arguments, in specific cases with $F_{\mu\nu} \ne 0$
singularity formation has been shown by Clayton \cite{Clayton:2001vy}.

It is for these reasons that the Einstein-\AE ther literature does not
consider the Maxwell vector kinetic term
\cite{Jacobson:2008aj}. Interestingly there is little rigorous
understanding for what choices of parameters $c_i$ do give well
behaved vector dynamics.  Clearly a caustic singularity is signalled
by the divergence of the vector field becoming infinite. Hence it is
expected that by adding the term $c_2$ appropriately, which directly
energetically weights this divergence, one can dynamically suppress
singularities. For example, taking the Maxwell case together with the
additional term $c_2$, we obtain a vector equation of motion,
\begin{equation}
  (\delta^{\alpha}_{\mu} + A^\alpha A_\mu) \left[\partial_\nu F^{\nu \mu} + c_2 \partial^\mu (\partial\cdot A)\right]=0.
\end{equation}
We note that when $c_2=1$ the equation of motion is essentially the
wave equation, and hence in a regular geometry we would certainly not
expect singular behaviour. How large $c_2$ should be to avoid
singularities is an interesting open problem.

\begin{figure}[t]
\begin{center}
\includegraphics[angle=0,width=2.9in]{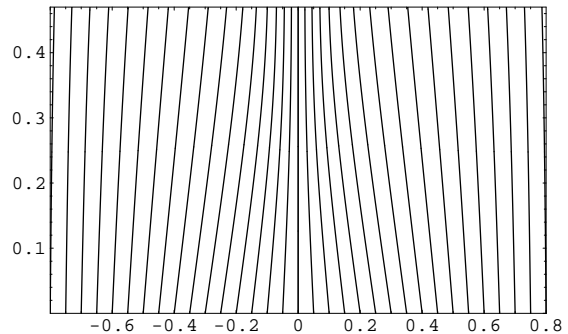}
\caption{\emph{Flat Space} Same as figure~\ref{flatplot1} but for
  $c_2=0.2$ i.e. a divergence term included in the kinetic part of the
  Einstein-\AE ther vector action. The addition of a divergence term
  suppresses the formation of caustics.}
\label{flatplot2}
\end{center}
\end{figure}

Ignoring gravity and matter, we simply plot a vector solution to the
above equation in 1+1 flat space in figures\ref{flatplot1}
and~\ref{flatplot2}. The left frame is for $c_2=0$, the right for $c_2
= 0.2$, and both have the same initial data which satisfies
$F_{\mu\nu}=0$. Note that the right frame cannot be lifted simply to a
solution of Einstein-\AE ther since for $c_2 = 0.2$, $F_{\mu\nu}$ will
not remain zero, and the vector necessarily contributes to the
gravitational stress tensor. However, clearly by eye we see a change
in behaviour, where the small amount of positive $c_2$ avoids the
caustic, leading to an asymptotic vector solution which is aligned
with time.

\section{Vector field dynamics in the absence of matter}\label{analyticshock}

Whilst in Einstein-\AE ther we can exhibit the large class of
solutions of the vector field in (\ref{AEsoln2}) which lead to caustic
singularities, it is far from obvious that the same occurs in
TeVeS. For these solutions, $F_{\mu\nu}=0$ and this leads
geometrically to the integral curves of the vector simply being
geodesics that fall into the gravitational wells created by the
matter. However in TeVeS there is direct coupling between the vector
and matter, and hence in the presence of matter one can not have
$F_{\mu\nu}=0$. If $F_{\mu\nu}\ne0$, then the curves of the vector do
not follow geodesics, and we cannot argue that they must cross forming
caustic singularities, even if we suspect they might.

However we can make some analytic progress in the absence of matter so
$\tilde T_{\mu\nu} = 0$. Then we may consistently truncate to
solutions with constant scalar field. Initial data with constant
scalar and vanishing scalar time derivative, i.e. $\partial_\mu \phi =
0$ on an initial Cauchy surface, evolves to have constant scalar.

We stated earlier that we are interested in the strong acceleration
regime, for example we look at the dynamics on solar system scales or
smaller. We denote this scale of interest by $L$. Since the acceleration regime is actually determined
by the scalar gradient, a precisely constant scalar corresponds to
exactly the opposite, the low acceleration MOND regime, even if all
other dynamical fields have characteristic scales given by $L$. However, we are envisaging a physical situation in which the TeVeS scalar has long wavelength fluctuations set by surrounding matter, for example set by the galaxy in which the region of interest is embedded. These fluctuations will be taken to have gradient large enough to place our region of interest into the Newtonian regime, which translates to the condition that the scalar should vary on lengths set by the TeVeS scale $\ell$. In Appendix \ref{argument}, we show that given this vast separation of scales, $L\ll \ell$, a solution to the TeVeS equations where we instead take exactly constant scalar, ignore $\mathcal{F}$ and set $\mu=1$, is a good approximation to the full TeVeS equations, within the scale of interest $L$. \footnote{It is interesting that the analytic
  analysis here does actually hold for exactly constant scalar - the
  function $\mathcal{F}$ vanishes in this limit. However that is not
  the physical regime we are interested in. In any physical context
  there will always be some small scalar gradients.}

For a constant scalar the vector equation reduces to
\begin{equation} 
K \nabla_\beta F^{\beta\alpha} +\lambda A^\alpha = 0. 
\end{equation} 
Consider starting with initial data on a Cauchy surface $\Sigma$ at $t
= 0$ where $F_{\mu\nu}=0$. Since $A^\mu$ is timelike, the $t$
component $A^t$ cannot vanish, and hence the $t$ component of the
above vector equation sets $\lambda = 0$ on this initial data
surface. However, the remaining components $i = 1 \dots 3$ then
determine $\partial_t F^{t i} = 0$ at $t = 0$. Furthermore the Bianchi
identity for $F_{\mu\nu}$, $\nabla_{[\mu} F_{\nu\alpha]} = 0$ implies
that $\partial_t F_{ij} = 0$ at $t = 0$. Together these imply that
$\partial_t F_{\mu\nu} = 0$ and so $F_{\mu\nu}$ remains zero when
evolved off the surface $\Sigma$ . Hence we see that starting with
initial data $\partial_\mu \phi = F_{\mu\nu} = 0$ on $\Sigma$ implies
(in the absence of matter) that the scalar is constant, $F_{\mu\nu} =
0$ and $\lambda = 0$ for all $t$. As discussed above, for $F_{\mu\nu}
= 0$ the vector can then be written as $A_\mu = \partial_\mu \chi$
with the timelike constraint $A^\mu A_\mu = -1$ giving the p.d.e. in
equation (\ref{eq:gaugepde}) which can be used to evolve $\chi$.
Hence the initial data for the vector can be parametrized by the
function $\chi$ on $\Sigma$ which determines the direction of the
vector on $\Sigma$.

The dynamics of \teves\ in this truncation are the same as those of
Einstein-\AE ther with Maxwell kinetic term and no matter. Hence as we
claimed above, for $F_{\mu\nu} = 0$ where vector integral curves are
timelike geodesics, we should expect to be able to form
caustics. While this is true, and indeed figure~\ref{flatplot1} gives
an example in 1+1 for the Minkowski geometry, there is no matter to
focus the geodesics and hence it isn't obvious how generic caustic
formation would be. If we start with initial data in this class of
solution - i.e. suitable data for the metric, together with the
function $\chi$ which specifies initial data for the vector - are the
initial data that develop to a singularity generic, or a special case?
We now address this by precisely characterizing when initial data will
form a caustic. We note that while our analysis is given in the
context of TeVeS, precisely the same argument can be made in the
context of Einstein-\AE ther theory, although we know of no previous
literature doing so.

Hence we consider TeVeS in the absence of matter, with constant
scalar, and with $F_{\mu\nu}=0$. We note that the TeVeS vacuum, with
Minkowski geometry and $A^\mu = (\partial_t)^\mu$ is in this class,
and hence we may regard the class as a restricted (although not
necessarily small) deformation of the TeVeS vacuum. The equation for
the metric immediately gives $R_{\mu\nu} = 0$. Thus the class covers
gravity wave spacetimes and black hole exteriors (with constant
scalar).

We now briefly review some basic facts in GR. For a congruence of
timelike geodesics, parametrized by proper time $\tau$ with tangent
vector field $\xi^\mu$, with $\xi^\mu \xi_\mu = -1$, we may define a
tensor field
\begin{equation} 
B_{\mu\nu} = \nabla_\nu \xi_\mu, 
\end{equation} 
which then satisfies $B_{\mu\nu} \xi^\mu = B_{\nu\mu} \xi^\mu = 0$. We
define the expansion $\theta$, shear $\sigma_{\mu\nu}$, and twist
$\omega_{\mu\nu}$ as,
\begin{eqnarray}
\theta & = & B^{\mu\nu} h_{\mu\nu}, \nonumber \\
\sigma_{\mu\nu} & = & \frac{1}{2} B_{(\mu\nu)} - \frac{1}{3} \theta h_{\mu\nu}, \nonumber \\
\omega_{\mu\nu} & = & \frac{1}{2} B_{[\mu\nu]},
\end{eqnarray}
where $h_{\mu\nu} = g_{\mu\nu} + \xi_\mu \xi_\nu$ is the projector
onto the tangent space orthogonal to the timelike geodesics. Then
Raychaudhuri's equation is,
\begin{equation}
\frac{d\theta}{d\tau}=-\frac{1}{3}\theta^2 - \sigma_{\mu\nu}\sigma^{\mu\nu} - \omega_{\mu\nu} \omega^{\mu\nu} - R_{\mu\nu}\xi^{\mu}\xi^{\nu} .
\end{equation}
Now we consider applying this result to our situation. Recall that
since our solutions have $F_{\mu\nu} = 0$, then the integral curves of
$A^\mu$ are timelike geodesics, and moreover the tangent vector
$A^\mu$ has unit norm. Hence we may take the $\xi^\mu$ above to be
$A^\mu$. Then since $F_{\mu\nu} = 0$ the twist $\omega_{\mu\nu}$
vanishes, and hence the congruence is hypersurface
orthogonal. Furthermore we have $R_{\mu\nu} = 0$, and using the fact
that $\sigma_{\mu\nu}\sigma^{\mu\nu} \ge 0$ therefore arrive at the
expression,
\begin{equation}
\frac{d\theta}{d\tau} \le -\frac{1}{3}\theta^2, 
\end{equation}
where $\theta = ( \nabla_\mu A_\nu ) ( g^{\mu\nu} + A^\mu A^\nu) =
\nabla \cdot A$ since $A^\mu A_\mu = -1$. Thus we have the result,
\begin{equation}
(\nabla\cdot A)^{-1}(\tau)\geq (\nabla\cdot A)^{-1}_0 + \frac{1}{3} \tau,
\end{equation}
along a geodesic with $\nabla\cdot A = (\nabla\cdot A)_0$ at the point
where the geodesic intersects the initial Cauchy surface $\Sigma$. So
we conclude that if $\nabla\cdot A<0$ anywhere on the initial
hypersurface $\sigma$, within a proper time $-3(\nabla\cdot
A)^{-1}_0$, $\nabla\cdot A$ diverges, signaling that the geodesic
congruence ends at a caustic singularity.

In summary, we have obtained the following result: In the absence of
matter smooth initial data with $\partial_\mu \phi = F_{\mu\nu}=0$ on
a spacelike hypersurface $\Sigma$ will evolve to form a caustic
singularity \emph{if $\nabla\cdot A< 0$ anywhere on $\Sigma$.} Note that while these are solutions with exactly constant scalar, the timescale of caustic formation is set by $L$ and so they will still be good approximations to the TeVeS equations in the strong acceleration regime, as discussed in Appendix~\ref{argument}.

Whilst this class of solutions is clearly restricted it is still
physically reasonable and in particular includes initial data close to
the TeVeS vacuum. The initial data includes the initial data for the
metric and for the vector, the function $\chi$ on $\Sigma$. The
condition that $\nabla\cdot A< 0$ at any point on the initial data
surface is very weak, and certainly generic within our restricted
class.  For example, consider the small perturbation from the \teves\
vacuum where the metric is taken to be Minkowski and the vector near
the initial surface $\Sigma$ at $t = 0$ is given by $\chi = -t +
\delta \chi$, for small $\delta \chi$. Then the singularity condition
on $\Sigma$, $(\nabla\cdot A) \simeq \nabla_i^2 \delta\chi < 0$ will
be generically satisfied in the region surrounding a maximum in $\chi$
on $\Sigma$.

Hence our result very clearly highlights the fact that caustic
singularities do indeed occur in TeVeS. Note that we have bounded the
time to form the singularity by the initial data - no dimensional
constants enter that might be used to tune away the singularity. It is
worth emphasizing that if we were considering a gauge theory where
$A_\mu$ were a gauge potential, then such caustic singularities would
not be a concern. One could simply gauge them away. However, the
Lagrange multiplier and coupling of the vector to the matter here mean
there is no such gauge symmetry, and a singularity of the type here is
a true physical singularity. The classical evolution of the theory is
ill-posed once the singularity has formed.

In a sense we are done. Whilst in the absence of matter for more
general initial data we have no argument to show caustics form, it
certainly seems likely. This fact is already sufficient to render the
dynamics of the TeVeS theory rather dubious. However since
phenomenologically it is the interaction with matter that is of
tantamount interest in TeVeS it is still interesting to understand
whether matter ameliorates the situation or instead makes it worse and
we spend the remainder of the paper investigating this.

An interesting and possibly related topic is the linear instability
noted by Seifert~\cite{Seifert:2007fr} when considering perturbations
of spherically symmetric backgrounds in Einstein-\AE ther and \teves\
using the methods of~\cite{Seifert:2006kv}. Clearly the phenomenon of
caustic singularity formation is essentially non-linear and one would
not expect to see it in linear theory. However the dynamics of such
singularity formation may be associated with a growing unstable linear
mode for the vector field, and it would be interesting to investigate
whether there is indeed a link between our discussion and Seifert's
linear instability. Seifert has discussed modifying \teves\ in the
same sense we do later, and his work led to Skordis considering the
same modification we present later in the context of cosmological
perturbation theory.

\section{Vector field dynamics and new static black hole geometries}\label{bh}

In this section we find new classes of black holes in TeVeS where the
Einstein metric is static, but the vector field has non-trivial
dynamics, and its integral curves fall through the horizon. The
previous static black hole geometries of Giannios
\cite{Giannios:2005es} and of Sagi \& Bekenstein \cite{Sagi:2007hb}
have had the vector aligned with the killing time, which naively
suggests the vector field in the exterior regions to slowly moving
matter might dynamically wish to align with the matter's natural
frame. However, the existence of our new solutions clearly shows that
this is not to be expected, and that the dynamics of the vector
exterior to a region containing matter may be very complicated, and in
particular is likely to want to fall towards the matter and may form
caustic singularities. For an actual matter source rather than a black
hole where the integral curves cannot disappear through a horizon,
accumulation of curves as the vector falls towards the matter are more
likely to form singularities. Indeed in later sections our simulations
with matter show that this is the case.

Later in the paper we will suggest a modification of the TeVeS theory
to avoid caustic singularity formation. It is worth noting that the
black hole solutions presented below will \emph{not} be solutions in
this modified theory and hence we avoid going into detailed
phenomenology for these solution here. It would be interesting to
study black hole solutions in the modified theory we suggest, and we
make some comments on this in the concluding discussion section
\ref{disc}.

%
%

As discussed in the previous section we will be interested in length scales $L\ll \ell$, where the TeVeS scalar is approximately constant in the region of interest but varies enough to place the region in the strong acceleration regime. Thus, as discussed in Appendix~\ref{argument}, we may consider solutions to the TeVeS equations with exactly constant scalar, ignoring $\mathcal{F}$ and setting $\mu=1$ as a good approximation.

\subsection{New static black hole geometries with constant scalar}\label{constphisol}

It was shown by Giannios in \cite{Giannios:2005es} that if the vector field $A$ is
aligned with the time translation Killing vector, then the solution
for $\phi$ is singular unless it is constant. Letting that constant be $\phi_c$, the solutions take the form,  
\begin{eqnarray} 
\phi&=&\phi_c, \\
A^\mu&=&\left(\frac{1}{\sqrt{T(r)}},0,0,0\right),\label{vectorcomps0}\\
ds^2&=&-T(r)dt^2+R(r)dr^2+r^2\left(d\theta^2+\sin^2\theta d\varphi^2
\right),\nonumber\\
\,
\end{eqnarray}
For our new solutions we again have a static Einstein metric and constant scalar, but now take $A^r \ne 0$,
\begin{equation} 
A^\mu=\left(A^t(r),A^r(r),0,0\right),\label{metricansatz0}
\end{equation}
The scalar field equation is
trivially satisfied for a vacuum spacetime. The vector field equation becomes
\begin{equation} K\nabla_\beta F^{\beta \alpha}+\lambda
  A^{\alpha}=0.\label{vecbody}.
\end{equation}
For the case where $\alpha=r$ the first term
vanishes leaving
\begin{equation}
\lambda A^r=0,
\end{equation}
and thus for $A^r\not=0$ we have $\lambda=0$. In this case the field
equations become those of Einstein-Maxwell theory for a particular
choice of gauge. Given this we expect to find Reissner-Nordstr\"om
(RN) black holes, and one can check this is indeed the general
solution - we give the argument in appendix \ref{constbhdetails}
\footnote{Note that in \cite{Sagi:2007hb} the authors derive a RN
  solution in the MF by introducing a `true' Maxwell field. Their
  charge is therefore unrelated to the charge $Q$ in our
  solution.}. We find,
\begin{widetext}
\begin{eqnarray}
g&=&\mbox{Diag}\left[-\left(1-\frac{2M}{r}+\frac{Q^2}{r^2}\right),\left(1-\frac{2M}{r}+\frac{Q^2}{r^2}\right)^{-1},r^2,r^2
\sin^2(\theta)\right], \label{gsol}\\ 
A_\mu&=& \partial_\mu \Phi + \delta^t_\mu \sqrt{\frac{2}{K}} \frac{Q}{r},\label{asol1}\\
\Phi & = & -t \pm \int dr \left(1-\frac{2M}{r}+\frac{Q^2}{r^2}\right)^{-1}\sqrt{\left(\frac{2}{K}-1\right)\frac{Q^2}{r^2}+\left(M+\sqrt{\frac{2}{K}}Q\right)\frac{2}{r}},\label{gaugechoice}
\end{eqnarray} 
\end{widetext}
and hence see that the solution is indeed simply RN where the gauge
freedom for the field $A$, specified by the function $\Phi$, has been
fixed up to a sign by the Lagrange multiplier constraint that $A$ has
unit timelike norm. Note that for $Q = 0$ this gauge transformation
$\Phi$ is the Lema\^itre time coordinate, i.e. the coordinate time
experienced by in-falling geodesic observers. Clearly, $F_{\mu\nu}=0$
when $Q=0$ and so the integral curves of $A$ are geodesics of the EF
geometry.

\begin{figure}[t]
\begin{center}
\includegraphics[angle=0,width=3.0in]{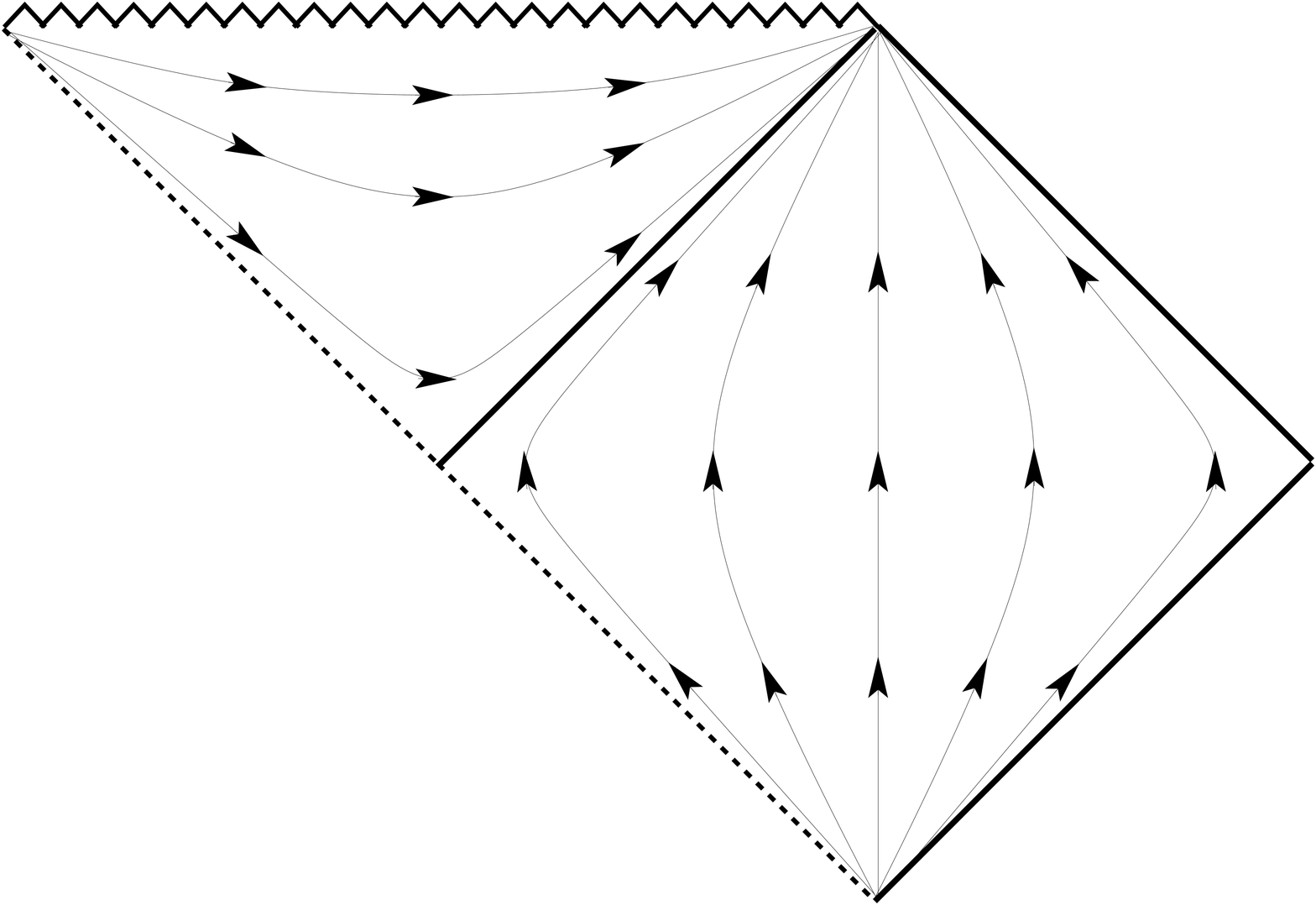}
\caption{Penrose Diagram for the Schwarzschild MF
  Spacetime where the vector field is aligned with the Killing time as
  in the Giannios solutions~\cite{Giannios:2005es}.}
\label{schwarzschild1}
\end{center}
\end{figure}

One may show that the matter frame metric is RN too, after an
appropriate coordinate transformation. We cast $\tilde{g}$ into
Schwarzschild-type coordinates, $(\tau,\rho)$ defined as,
\begin{equation}
t(\tau,\rho)=e^{-\phi_c} \tau + f(\rho)  \qquad r(\rho)=e^{\phi_c}\rho,
\end{equation}
where $f$ is given by,
\begin{equation}
f_{,\rho}=-e^{\phi_c} \frac{\tilde{g}_{rt}}{\tilde{g}_{tt}}.
\end{equation}
These transformations leave all components $\tau$ independent, and
$\tilde{g}$ in standard RN form in Schwarzschild coordinates
\begin{equation}
  \tilde{g}_{\tau\tau}= -\frac{1}{\tilde{g}_{\rho\rho}} = -\left(1-\frac{2\tilde{M}}{\rho} + \frac{\tilde{Q}^2}{\rho^2}\right) \qquad \tilde{g}_{\tau\rho}=0,
\end{equation}
where now the mass and charge are given as,
\begin{eqnarray}
e^{\phi_c} \tilde{M}&=&e^{-4\phi_c} M - \left(1-e^{-4\phi_c}\right)\sqrt{\frac{2}{K}} Q,\\
e^{2\phi_c} \tilde{Q}^2 &=& \left(e^{-4\phi_c} +\left(1-e^{-4\phi_c}\right)\frac{2}{K}\right)Q^2. 
\end{eqnarray}
This shows the metric is still RN, but with shifted horizon positions.
For example, in the Schwarzschild case ($Q=0$) we have that the MF and
EF horizons are related by
\begin{equation}
r_{\mathcal{H}(\tilde{g})}=e^{-4\phi_c}r_{\mathcal{H}(g)}.\label{horizons}
\end{equation} 
Bekenstein \cite{Bekenstein:2004ne} demonstrates the speed of
scalar perturbations (at fixed background vector) and vector
perturbations (at fixed background scalar) is sub-luminal with respect
to electromagnetic propagation only if $\phi>0$ everywhere. So we
conclude that in this case, the horizon seen by the matter fields is
smaller than the horizon for the gravitational fields ($\phi,A,g$).

\begin{figure}[t]
\begin{center}
\includegraphics[angle=0,width=3.0in]{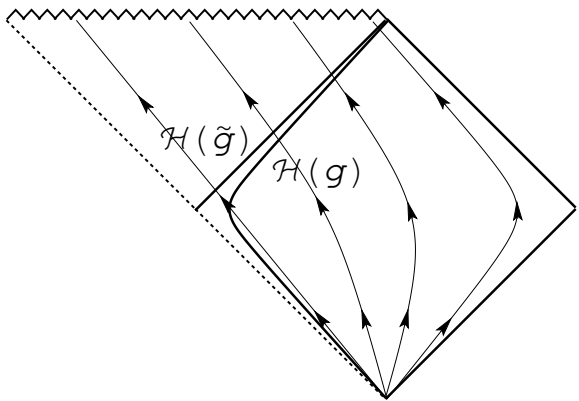}
\caption{The Penrose Diagram for the generalized Schwarzschild MF
  Spacetime ($Q=0$ case) with the EF horizon position. In this case
  the vector field is free-falling along geodesics of $g$ through both
  horizons.}
\label{schwarzschild2}
\end{center}
\end{figure}

We now approach TeVeS solutions from another direction. Starting with
the standard RN solution in Einstein-Maxwell theory, we may obtain a
vacuum solution of TeVeS with constant scalar and $\lambda = 0$
provided we may choose a gauge such that the vector potential
satisfies the TeVeS constraint $A^\mu A_\mu = -1$. Hence we find a
large class of solutions with metric (\ref{gsol}) and vector of the form (\ref{asol1}) but where, as in the earlier equation \eqref{eq:gaugepde}, we regard the gauge condition as a first order p.d.e. in time for $\Phi$,
\begin{eqnarray}
\partial_t \Phi &=& - \sqrt{\frac{2}{K}} \frac{Q}{r} + \frac{1}{(- g^{tt})}\Bigl(  g^{ti} \partial_i \Phi -\Bigr. \nonumber\\
 &&\left. \sqrt{ (- g^{tt}) ( 1 + \partial^i \Phi \partial_i \Phi) + ( g^{ti} \partial_i \Phi)^2} \right).
\label{eq:gaugepde2}
\end{eqnarray}
Hence we may take the solutions to be characterized by the charges $M,
Q$ and also $\Phi(t = 0)$, from which we evolve to construct $\Phi$
for all $t$. The solution above in equation \eqref{gaugechoice} is a
stationary solution to this p.d.e. for $\Phi$. In general however the
solutions to this p.d.e. will have complicated time dependence.

Note that for $Q = 0$, this class of solutions has $R_{\mu\nu} =
F_{\mu\nu} = 0$ as well as constant scalar and hence falls into our
earlier class discussed in section \ref{analyticshock}. Thus in this
black hole background we can again precisely characterize caustic
singularity formation by the previous statement that, letting $(\nabla
\cdot A)_0$ be the value of $\nabla \cdot A$ on a surface $\Sigma$,
then if $(\nabla \cdot A)_0 < 0$ at any point on $\Sigma$ a
singularity will form within a proper time $-3 (\nabla \cdot
A)^{-1}_0$ measured along the future of the integral curve of $A$
through that point.  Since this condition is generic within our class
of solutions for $Q = 0$, we see that caustic singularity formation is
to be expected in the exterior of these black holes. Whilst we have no
argument that the same is true for $Q \ne 0$, we expect it is likely.

The presence of the matter appears to attract the vector field
integral curves. Naively this focusing would seemingly make caustic
singularity formation more likely. However, interestingly the presence
of the horizon actually renders the singularity formation less severe
in the sense that if the time to singularity formation is sufficient
that the integral curve of $A$ has fallen inside the horizon, an
external observer need not care. Indeed in the stationary solution
\eqref{gaugechoice} above this precisely happens, with a caustic
singularity occurring at the black hole singularity itself. Of course,
if the matter source is not a black hole, but rather a compact object
without horizon, then we still expect the vector field curve
attraction, but now there is nowhere for the curves to go, and hence
the expectation that the matter focuses the vector to form
singularities would hold. Later in the paper we investigate this.

We conclude this section by commenting that we have examined the case
of constant scalar, stationary black hole solutions. One might wonder whether stationary 
solutions with non-constant scalar can be found, which share the symmetry and asymptotics of those found here. We address this
questions in the Appendix \ref{singularbhdetails}, finding evidence
that no solutions exist near to the ones above with constant
scalar. We show that for a linear perturbation of the scalar about our
constant scalar solution a singular develops exterior to the horizon,
and that performing a full non-linear numerical evaluation of such a
solution one finds both the EF and MF metrics are nakedly singular. We
cannot argue that no non-constant scalar black holes exist
\footnote{Giannios \cite{Giannios:2005es} and more recently, Sagi \&
  Bekenstein \cite{Sagi:2007hb} have found black hole solutions where
  the scalar diverges logarithmically, but the MF metric remains
  regular. Note that these solutions are not `near' ours in the sense
  that the vector field has a very different behaviour as we have
  discussed in this section.} but do expect there are none that are
qualitatively similar to the constant scalar solutions we have found. Note that this is compatable with the argument presented in Appendix~\ref{argument} since the kind of non-trivial scalar considered there need not be static nor share the same asymptotics or symmetry.

\section{Vector field dynamics and matter: Numerical simulation}
\label{shockformation}

In section \ref{analyticshock} we have shown that in the absence of
matter, a large class of deformations of the TeVeS vacuum initial data
quickly terminate in caustic singularities upon time evolution. In
section \ref{bh} we have shown that black hole solutions of TeVeS have
complicated vector dynamics, which again include caustic
singularities, and in particular that the black hole appears to
attract the integral curves of the vector field toward it. We might
then expect this to occur for any matter source, and then imagine that
such an attraction which focuses the integral curves is likely to
generate caustic singularities. This is too quick however, as matter
couples to the vector field in TeVeS in a complicated fashion, and
hence we have little intuition or analytic control over what happens.

It is the subject of this section to investigate the vector dynamics
in the presence of matter using full numerical evolution of the TeVeS
equations of motion. To make progress we restrict ourselves to
spherical symmetry. We are then able to consider both gravitational
collapse of a matter scalar field, and evolution of an initially near
static boson star. In both cases we find the vector curves in the
region exterior to the matter are indeed attracted towards the matter
and do form caustic singularities.  One might be concerned that
imposing spherical symmetry restricts to a rather special class of
solutions which focuses energy at the origin of spherical
symmetry. However the caustic singularities actually form away from
the origin and hence the singularities themselves locally have a
planar symmetry, and seem not to result from the peculiarities of
spherical symmetry. In both cases the TeVeS scalar is fully dynamical and non-constant, hence justifying the Newtonian regime approximation ($\mu=1$, neglecting $\mathcal{F}$). It remains smooth where the caustics form indicating that the scalar plays no role in the pathalogical vector dynamics.

Full details of the numerical implementation and convergence and
constraint tests are postponed to Appendices \ref{numericalmethod} and
\ref{constrainttesting}.

\subsection{Scalar field collapse}\label{bh2}

Our first matter system is the collapse of a complex scalar field. We
perform an integration of the field equations from an initial
spherical shell of scalar matter. We use a canonical complex scalar field
$\chi$, whose action is constructed using the matter frame metric
$\tilde{g}$. Using time symmetric initial data and a radial gaussian
shell for the matter field $\chi$, the matter energy density will
split into ingoing and outgoing components and for sufficient
amplitude of the initial shell, the ingoing component will be focused
at the origin into a high enough energy density to form a black
hole. We use the coordinate system
\begin{equation}\label{bhcoords}
ds^2=-T^2(t,r)dt^2+e^{R(t,r)}dr^2+r^2d\Omega_2^2.  
\end{equation} 
which clearly covers only the exterior region of any black hole that
might form. We emphasize that since the matter $\chi$ couples to the
TeVeS vector and scalar, both must be evolved and have non-trivial
dynamics.

For the metric we choose time symmetric initial data that satisfies
the constraint equations. The TeVeS scalar we take to be constant
initially, and time symmetric, and for the vector, we take $A^r = 0$
at $t = 0$ and choose $\dot{A}^r$ such that $F_{\mu\nu} = 0$ at $t
=0$. However, since there is matter, in contrast to the discussion in
the previous sections above, $F_{\mu\nu}$ will immediately evolve to
be non-zero, and $\phi$ to be non-constant.

\begin{figure*}[th]
\begin{center}
\includegraphics[angle=0,width=7.0in]{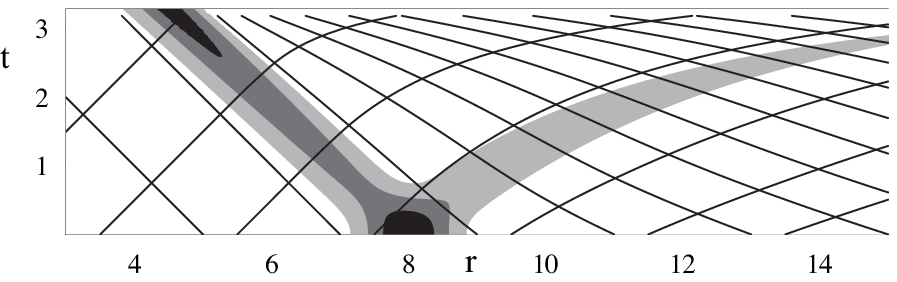}
\includegraphics[angle=0,width=7.0in]{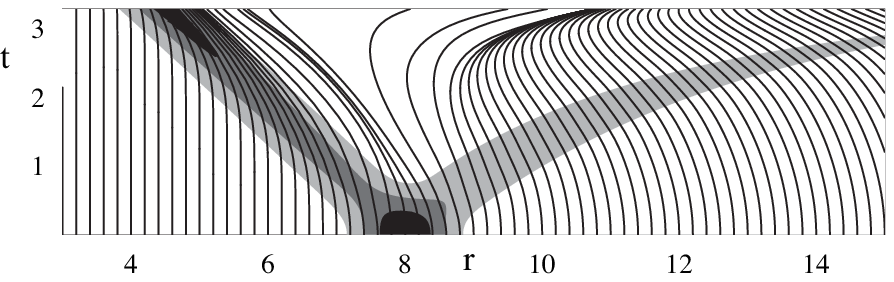}
\caption{Massless scalar field shell-collapse in an attempt to form a
  black hole.  \emph{Upper Frame:} The light cone structure is shown
  overlaying contours for the amplitude of the $\chi$ matter
  field. \emph{Lower Frame:} Integral curves for the TeVeS vector
  $A$. In this evolution we see a caustic singularity develop around
  $r \simeq 11$ at $t \simeq 3.2$. The light cone structure indicates
  an apparent horizon is likely to have formed if the evolution had
  not been terminated by the caustic. The initial data for the vector was $A_r = 0$ with $\dot{A}_r$ chosen so $F_{\mu\nu} = 0$. The coordinate system is
  (\ref{bhcoords}), with parameters $K=\kappa=0.01$}
\label{bh2fig}
\end{center}
\end{figure*}

\begin{figure*}[th]
\begin{center}
\includegraphics[angle=0,width=5.0in]{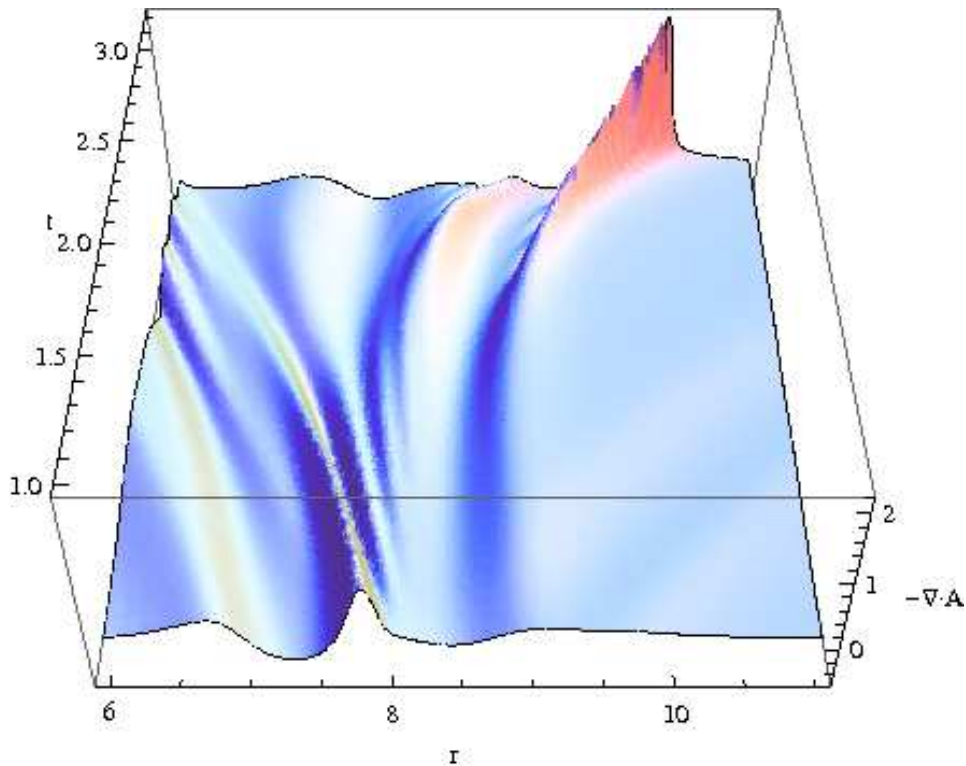}
\caption{$-\nabla\cdot A$ for the simulation of massless scalar field
  shell-collapse, as in figure~\ref{bh2fig}. Spiking of the
  4-divergence of $-A$ corresponding to the convergence of its
  integral curves is easily seen, signalling the formation of a
  caustic.}
\label{bhgradAfig}
\end{center}
\end{figure*}

\begin{figure*}[th]
\begin{center}
\includegraphics[angle=0,width=5.0in]{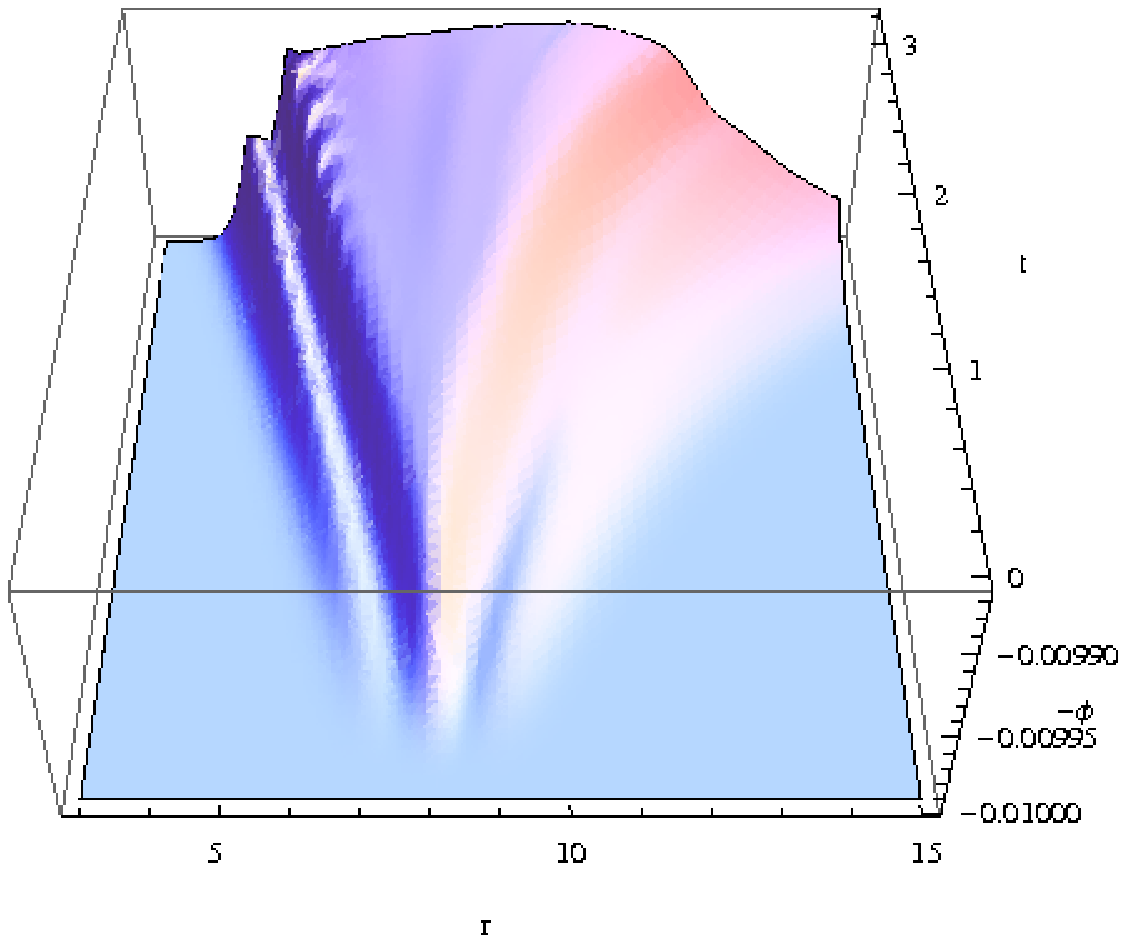}
\caption{$-\phi$ for the simulation of massless scalar field
  shell-collapse, as in figures~\ref{bh2fig} and~\ref{bhgradAfig}. The scalar field remains small and smooth up to the formation of the caustic.}
\label{bhphifig}
\end{center}
\end{figure*}

We find that for different initial Gaussian shells of $\chi$, of
sufficient amplitude to ensure non-linear dynamics when the ingoing
pulse reaches the origin (for weak amplitudes the energy density
simply passes through the origin and radiates to infinity as is the
case in usual standard gravity coupled to a scalar) rather similar
qualitative behaviour results. Figures \ref{bh2fig}, \ref{bhgradAfig} and \ref{bhphifig} give the results of a representative evolution. Even
though $F_{\mu\nu} \ne 0$, we do see that the vector field curves are
initially attracted to the matter shell. Evolution proceeds with the
ingoing pulse deforming the geometry as for a usual scalar
collapse. However before a horizon can form - recall our coordinate
system only covers the black hole exterior - we see the formation of a
caustic singularity. This is signaled by the divergence of $A$ which
develops a growing spike on constant $t$ slices as seen in figure
\ref{bhgradAfig}. At this point, dynamical evolution is no longer well
defined. As noted above, the singularity forms well away from the
origin of spherical symmetry. Figure \ref{bhphifig} shows that the TeVeS scalar remains small and smooth up to this point, which suggests that dynamically it does very little, if anything, to prevent singularity formation. It is interesting that whilst we have
found many candidate black hole end states for such a collapse in the
previous section, the actual dynamics of the collapse is sufficiently
badly behaved that we cannot even see an apparent horizon form.

\begin{figure*}[th]
\begin{center}
\includegraphics[angle=0,width=7.0in]{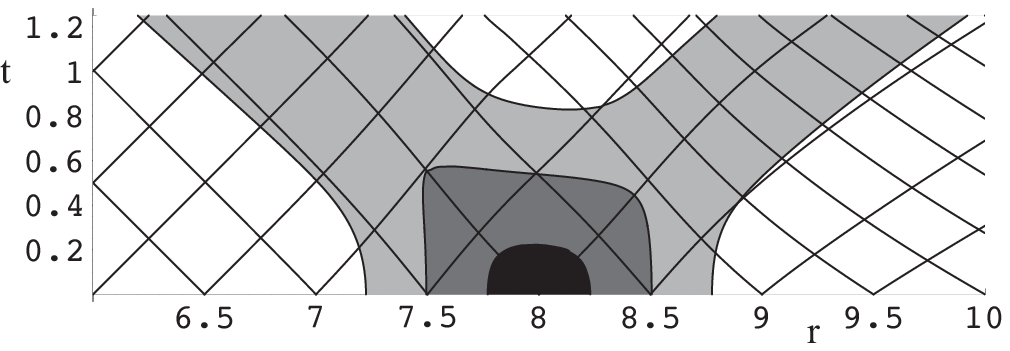}
\includegraphics[angle=0,width=7.0in]{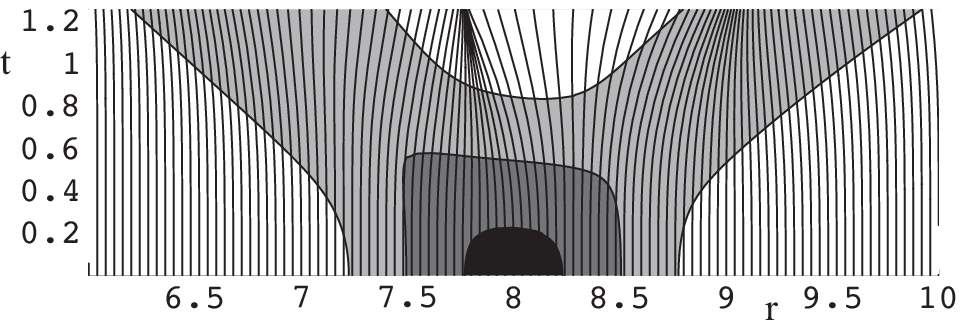}
\caption{ 
As for figure \ref{bh2fig}, although now with initial data $\dot{A}_r = A_r = 0$, so that $F_{\mu\nu} \ne 0$ initially. This simulation was performed with larger $K=\kappa=0.1$. We see again caustic formation, now much sooner.
}
\label{bh2fig2}
\end{center}
\end{figure*}

\begin{figure*}[th]
\begin{center}
\includegraphics[angle=0,width=5.0in]{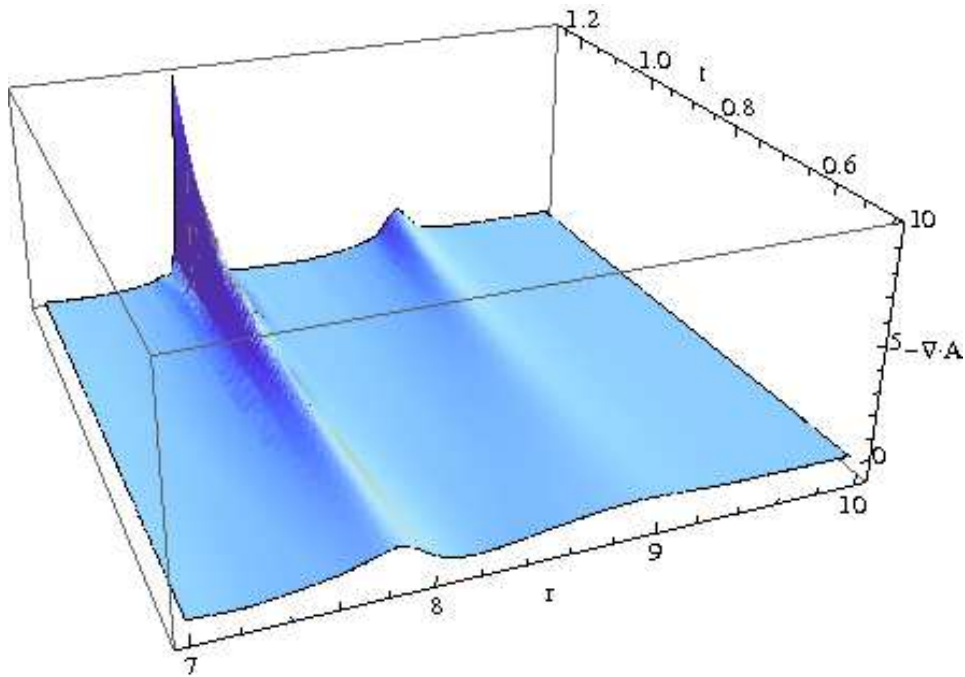}
\caption{$-\nabla\cdot A$ for the simulation in figure~\ref{bh2fig2}. }
\label{bhgradAfig2}
\end{center}
\end{figure*}

\begin{figure*}[th]
\begin{center}
\includegraphics[angle=0,width=5.0in]{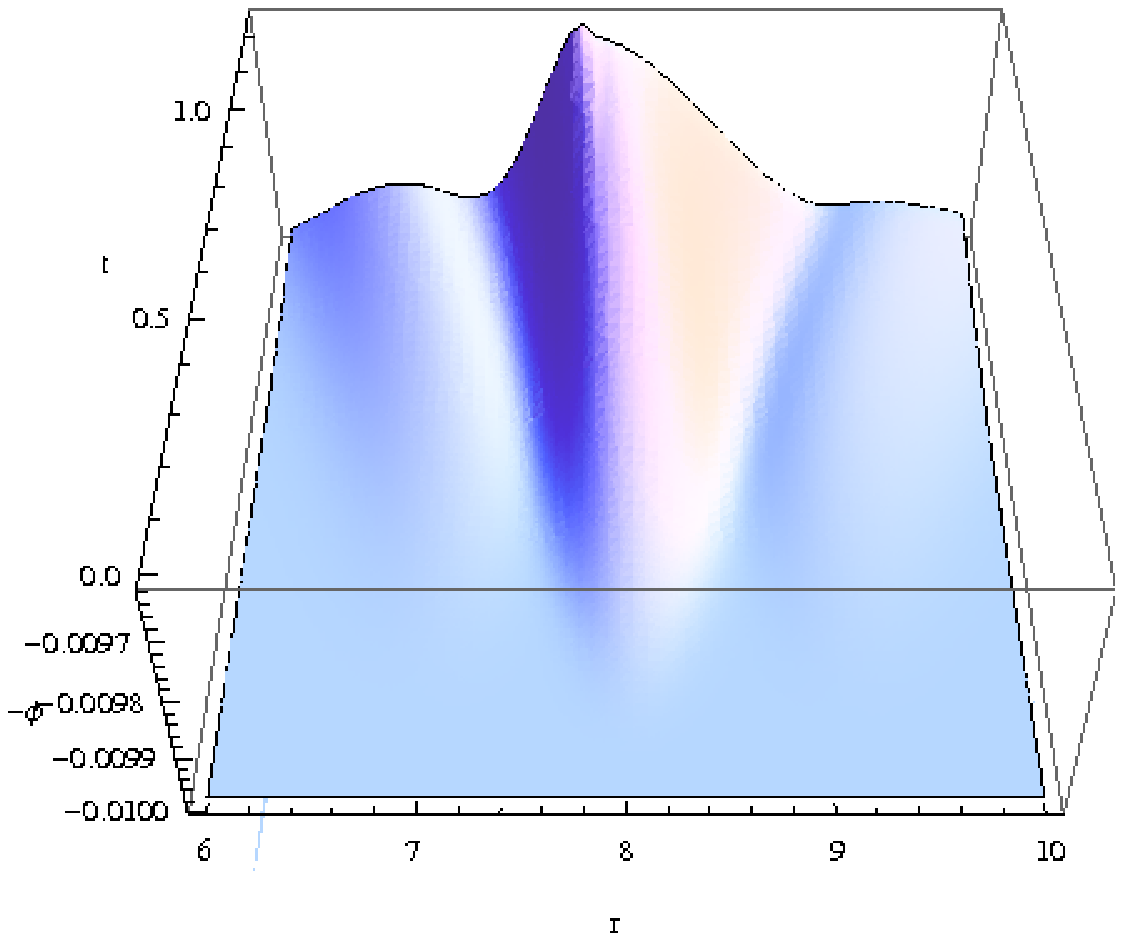}
\caption{$-\phi$ for the simulation in figures~\ref{bh2fig2} and~\ref{bhgradAfig2}.}
\label{bhphifig2}
\end{center}
\end{figure*}

One might worry that the choice of initial data with $F_{\mu\nu} = 0$
is somewhat special (even though $F_{\mu\nu}$ evolves to be non-zero).
\footnote{We thank Bekenstein for emphasizing this in private
  communication.}  Indeed, since our earlier analytic arguments were
for $F_{\mu\nu} = 0$ it is useful to check that caustics may also form
for initial data with $F_{\mu\nu} \ne 0$.  Another question is whether
the magnitude of $K, \kappa$ play a role in the singularity
formation. For these reasons we present the result of another
simulation in figures \ref{bh2fig2}, \ref{bhgradAfig2} and \ref{bhphifig2}. These
simulations used initial data for the vector where $\dot{A}_r = A_r =
0$. Hence the initial data as a whole is now time symmetric, and
$F_{\mu\nu} \ne 0$ initially. For the simulation shown we have also
taken larger $K, \kappa$. We observe caustic formation again. Indeed
the singularity forms earlier. Experimentally we find that for larger
$K, \kappa$ a caustic forms earlier, which is to be expected as the
vector is more strongly coupled to the dynamics of the other
fields. Thus we see that for two very different choices of the vector
initial data caustic singularities result.

\subsection{Perturbations to a boson star}\label{star}

Scalar field collapse is an extreme dynamical process which is highly
relativistic. It is interesting to consider whether an initially
non-relativistic matter source also seeds an attraction of vector
curves and subsequent caustic singularity. To this end we examine the
full dynamics of TeVeS in the presence of an initially quasi-static
boson star \cite{PhysRev.187.1767}.

We use an identical numerical method and boundary conditions to
integrate the field equations as for the scalar collapse above. To create the boson star we follow Gleiser \cite{PhysRevD.38.2376}; as a matter source we use a complex scalar, $\chi$, now with potential
$V(|\chi|)$. We begin by finding a static
boson star solution. This is achieved by imposing the following
separable solution to the $\chi$ equation of motion with the potential
$V(|\chi|)=m^2 \bar{\chi} \chi$
\begin{equation} 
\chi(t,r)=\chi_0(r) e^{\imath  \omega t}, 
\end{equation} 
where $\chi_0$ is real. While we term this a `static' star, we note
that in fact $\chi$ has the above phase rotation, although all other
fields are indeed static. The metric functions $T, R$ are taken to
depend only on $r$, and likewise for the TeVeS scalar. The TeVeS
vector is chosen to be aligned with Killing time, so $A_\mu =
(-T(r),0,0,0)$. The radial profile for each of these functions is obtained via
a shooting method - we fix the value of $\chi_0(0)$ and $m$ so that
the resulting solution will have flattened out well before the
boundary of our numerical grid. We then fine tune the value of
$\omega$ to obtain the profile for $\chi_0$ with no nodes; this is the
ground state star. We also choose the parameters so that the start has
a low density compared to its radius and hence $T(0,r) \simeq 1$, so
the backreaction of the star is weak - it is non-relativistic. Note
that for this static star $F_{\mu\nu} \ne 0$, and the TeVeS scalar is non-constant.

To consider a dynamical perturbation of this static star we take
similar initial data to the scalar collapse. We take $T, R$, the TeVeS
scalar and the matter scalar to have initial data simply given by that
on a constant time slice of the static boson star solution
above. However we now take $A^r = 0$ and $F_{\mu\nu} = 0$ at $t =0$
(although again $F_{\mu\nu}$ will not evolve to remain zero due to the
boson star matter). Thus the vector is not now aligned with Killing
time and dynamics will ensue. Note however that since the stars
considered are in the non-relativistic regime, this perturbation to
the vector field initial data is small. Hence the evolution is a
non-relativistic process in its early stages .

\begin{figure*}[th]
\begin{center}
\includegraphics[angle=0,width=6.0in]{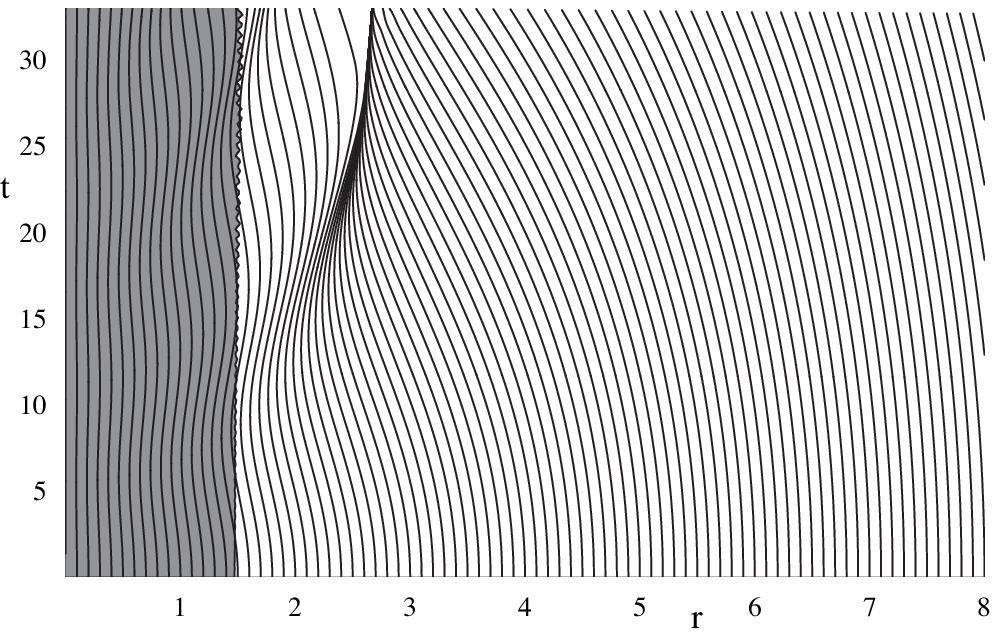}
\caption{A complex scalar field $\chi$ (grey shading) initially in the
  lowest static mode of a quadratic potential, as a matter source for
  a full TeVeS evolution. Shown also are the integral curves for the
  TeVeS vector $A$, illustrating the phenomenon of caustic formation
  \emph{where the initial data is quasi-static}.}
\label{starfig}
\end{center}
\end{figure*}

\begin{figure*}[th]
\begin{center}
\includegraphics[angle=0,width=5.0in]{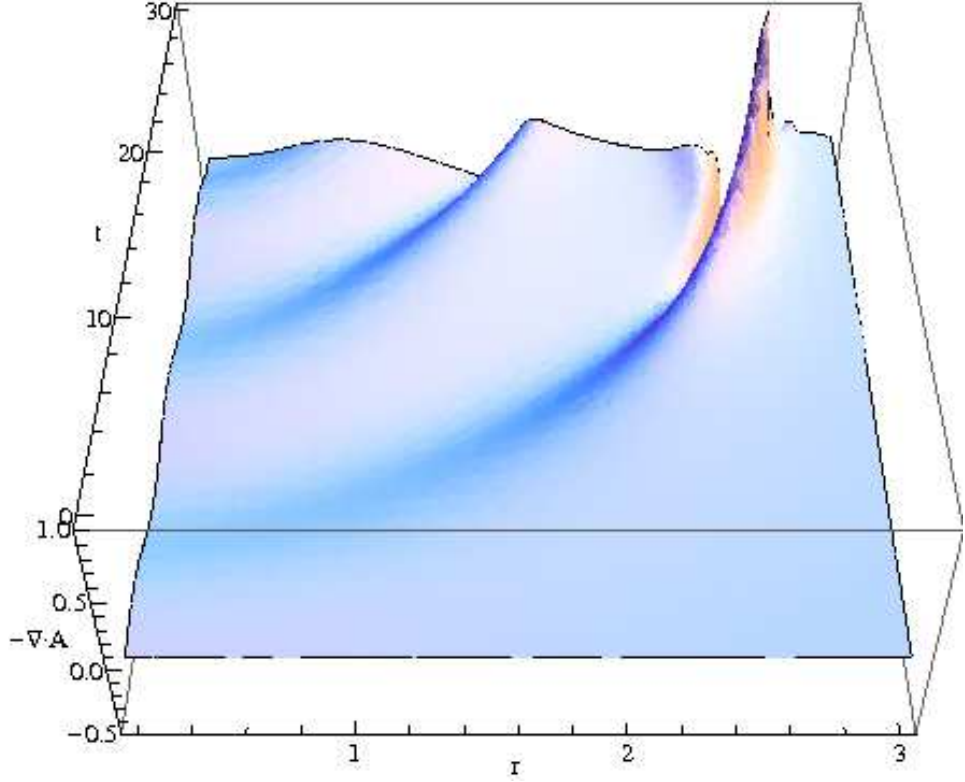}
\caption{$-\nabla\cdot A$ for the simulation of a complex scalar field
  as a ground state Boson star, as in figure~\ref{starfig}. Spiking of
  the 4-divergence of $-A$ corresponding to the convergence of its
  integral curves is easily seen, signalling the formation of a
  caustic singularity.}
\label{stargradAfig}
\end{center}
\end{figure*}

\begin{figure*}[th]
\begin{center}
\includegraphics[angle=0,width=5.0in]{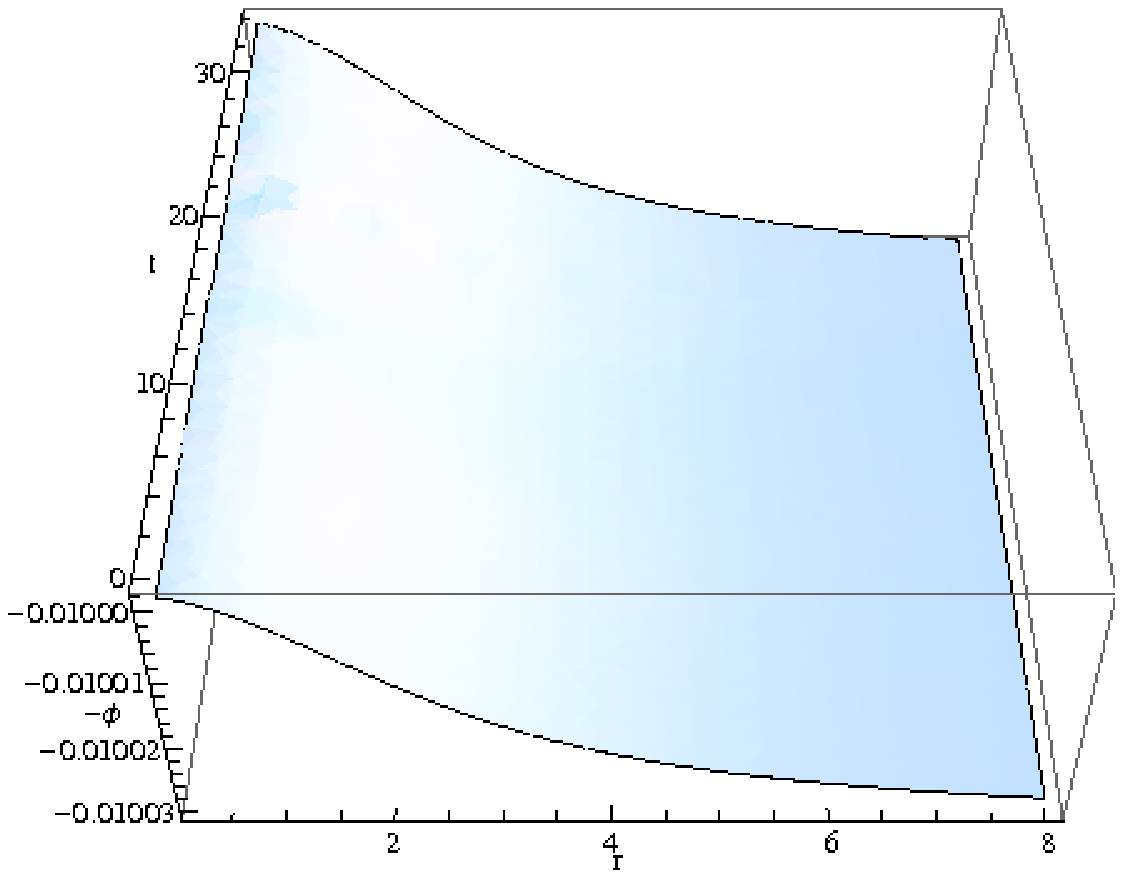}
\caption{$-\phi$ for the simulation of a complex scalar field
  as a ground state Boson star, as in figures~\ref{starfig} and~\ref{stargradAfig}.}
\label{starphifig}
\end{center}
\end{figure*}

We performed evolutions for a variety of star configurations,
obtaining qualitatively similar results. A representative choice is
illustrated in figure \ref{starfig}, \ref{stargradAfig} and \ref{starphifig}. This shows
that for our quasi-static initial configuration, the vector field
curves fall in towards the star and do evolve to form a caustic
singularity. However the singularity does not form in the interior of
the star as one might naively expect. This is essentially due to the
vector coupling to the matter, which apparently leads to a repulsive
effect as we see the integral curves are clearly repelled from the
origin of the spherical spatial geometry. While $F_{\mu\nu}$ is not
zero outside the star, as it is sourced by the stellar matter, and
then the region where $F_{\mu\nu}$ is non-zero propagates outward, we
see that it does not stop the vector curves falling towards the star
and eventually `colliding' with the curves that were `bounced' out of
the interior of the star. The singularity appears rather similar in
nature to that in the case of the scalar collapse, and figure
\ref{stargradAfig} clearly shows a growing vector divergence at a
finite radius as we approach the caustic. Figure \ref{starphifig} shows that once again the TeVeS scalar remains small and smooth up to the singularity.

Thus we have seen in this section that even starting with initial data
whose short term evolution is non-relativistic, pathological vector
behaviour may quickly follow. In particular we see visually that since
the integral curves are, at least initially, following an approximate
timelike geodesic, the timescale for this singularity formation is of
order the gravitational in-fall time. Thus in a Newtonian, quasi-static
regime such as the neighborhood of the Earth, one might expect caustic
singularity formation to occur on the order of hours, and after that
point classical evolution is ill defined. This poor dynamical
behaviour is clearly a serious obstruction to considering TeVeS as a
phenomenological theory of modified gravity.

\section{Modifying TeVeS to get a well behaved vector dynamics and a
  MoND limit}
\label{modsection}

Having demonstrated the formation of caustic singularities in TeVeS in
various contexts which render the classical dynamics of the theory
quickly ill defined, we now propose a correction to the vector part of
the action (\ref{tevesvectoraction}) which may ameliorate this
problem. The problem is essentially due to the Maxwell structure of
the vector action. There is no energy cost when the divergence of the
vector becomes large. Our modification is simply to introduce terms
that disfavour large divergences. We simply take the vector action to
be the most general diffeomorphism invariant action which is quadratic
in derivatives and consistent with the $A^2=-1$ constraint. This
action is of course precisely the one used for the vector in \AE $\,$
theory. We begin with
\begin{widetext} 
\begin{equation} S_{g+v} ={1\over 16\pi G}\int \left[R-
  \frac{K}{2}F_{\mu\nu}F^{\mu\nu} -
  \frac{c_{+}}{4}S_{\mu\nu}S^{\mu\nu} - c_2(\nabla_\mu A^{\mu})^2 -
  c_4\dot{A_\mu}\dot{A^\mu} +\lambda(A^2 +1)\right](-g)^{1/2} d^4
x,\label{modtevesvectoraction} 
\end{equation}
\end{widetext}
where $c_+ = c_1 + c_3$, $c_1 - c_3
= 2K$, $\dot{A}^\mu = A^\nu \nabla_\nu A^\mu$ and
$S_{\mu\nu}=\nabla_\mu A_\nu + \nabla_\nu A_\mu$. We retain the pure
TeVeS scalar action.  The metric redefinitions which may performed in
\AE $\,$ theory \cite{Foster:2005ec} to remove one of these terms are
no longer applicable here, as there are no non-trivial field
redefinitions which leave the scalar action form-invariant at the same
time as retaining the unit-norm constraint. As noted earlier, Seifert
\cite{Seifert:2007fr} has already proposed such a modification of
\teves\ motivated by finding a linear instability about certain
spherically symmetric backgrounds (those of Giannios
\cite{Giannios:2005es}) and leading from this Skordis
\cite{Skordis:2008pq} has recently derived the equations of motion and
studied the cosmological perturbation equations. Our emphasis here is
to check that the MoND limit is still recovered with this
modification, and this has not previously been addressed - without
this, of course, the modified \teves\ would be unlikely to provide any
alternative explanations for dark matter.

With this modified vector action, only the $\Theta$ term of the metric
field equation is affected, and we obtain \footnote{In pure TeVeS
  there are no connection terms appearing in the expression for
  $\Theta_{\mu\nu}$ above and so this does not introduce second
  derivatives of $A$ on the right hand side of Einstein's equations,
  and is advantageous for solving the initial-value problem. This is
  one reason Bekenstein gives for originally making this special
  choice of vector action. Unfortunately this is a luxury not afforded
  by the modified theory, which is not a problem in principle, but
  does complicate a treatment of the initial value problem. Hence we
  have not yet attempted to reproduce the numerical dynamical
  calculations of the previous section in the modified theory,
  although it would be interesting to do so.}
\begin{widetext}
\begin{eqnarray}
\Theta_{\mu\nu} = K\left(F_{\sigma\mu}F^{\sigma}_{\p{\sigma}\nu}-\frac{1}{4} F^2 g_{\mu\nu}\right) + \frac{c+}{2} \left( S_{\mu\sigma}S_{\nu}^{\p{\nu}\sigma} - \frac{1}{4} S^2 g_{\mu\nu} + \nabla_\sigma \left[  A^{\sigma} S_{\mu\nu}-S^{\sigma}_{\p{\sigma} (\mu} A_{\nu )} \right]\right)&&\nonumber\\
+ c_2 \left(g_{\mu\nu} \nabla_\sigma \left( A^\sigma \nabla \cdot A \right) - A_{(\mu}\nabla_{\nu)} \nabla \cdot A  - \frac{g_{\mu\nu}}{2} (\nabla\cdot A)^2 \right)&&\nonumber\\
 + c_4 \left(\dot{A}_\nu \dot{A}_\mu +  \dot{A}_\sigma A_{(\mu} \nabla_{\nu)} A^\sigma - \nabla_\sigma \left[\dot{A}^\sigma  A_\mu A_\nu \right]- \frac{g_{\mu\nu}}{2} \dot{A}_\sigma \dot{A}^\sigma \right)- \lambda A_\mu A_\nu,&&\nonumber\\
&&\nonumber\\
K\nabla_\mu F^{\mu\nu} + \frac{c_+}{2} \nabla_\mu S^{\mu\nu} + c_2 \nabla^\nu \left( \nabla\cdot A\right) - c_4 \dot{A}^\sigma \nabla^\nu A_\sigma + c_4 \nabla_\sigma \left(\dot{A}^\nu A^\sigma\right)&&\nonumber \\
+ \lambda A^\nu + 8\pi G \sigma^2 A^\mu \phi_{,\mu} g^{\nu\gamma} \phi_{,\gamma} = 8\pi G \left( 1-e^{-4\phi}\right) g^{\nu\mu} \tilde{T}_{\mu\gamma} A^{\gamma},&&
\end{eqnarray}
\end{widetext}
for the vector and scalar equations. We will refer to Bekenstein's theory as `pure TeVeS', and the theory
with the modification as `modified TeVeS'.

The new parameters introduced into the action, $c_+, c_2, c_4$ will
certainly be subject to physical constraints. The considerations are
likely to be similar to those constraining the Einstein-\AE ther
parameters, reviewed recently by Jacobson \cite{Jacobson:2008aj}. One
complication is that the physical fluctuation modes of the vector now
have different wave speeds when the more general vector action is
introduced. In particular this leads to new constraints from Cherenkov
radiation produced by cosmic rays \cite{Moore:2001bv,Elliott:2005va},
although since the matter couplings are different from those of
Einstein-\AE ther such analysis would likely have to be repeated for
the modified TeVeS. We leave determination of constraints on these
parameters for future work.

We have modified the TeVeS vector action in the hope that it will
alleviate the problem of singularity formation. An absence of caustic
formation is a crucial requirement for the theory to be dynamically
well behaved, though we do not attempt to assess whether this is
actually the case for our modified theory. We leave this as an
interesting open problem. Now, assuming that the dynamics of this
theory are in fact good, we would then require the theory to have the
appropriate phenomenology. That is, we would like modified TeVeS to
have inherited pure TeVeS's MoND limit. In order to check whether this
is the case, we perform a Newtonian analysis where the goal is to
obtain the equivalent of Poisson's equation for MF Newtonian
potential. We perturb the EF metric to leading order in the Newtonian
expansion as,
\begin{equation} 
g_{tt} = - 1- 2V  \qquad g_{ij} = (1-2V) \delta_{ij}  \qquad g_{ti} = 0,
\end{equation} 
where we linearize the equations in $V$, and ignore time derivatives
at leading order. The matter source has only the non-trivial component
$\tilde{T}_{tt} = \tilde{\rho}$. As for the EF metric, for the MF
metric we take,
\begin{equation} 
\tilde{g}_{tt} = - 1- 2 \Phi  \qquad g_{ij} = (1-2 \Phi) \delta_{ij}  \qquad g_{ti} = 0,
\end{equation} 
and in the Newtonian expansion only the time component of $A$ is
non-trivial at leading order, and is determined from $V$ by the
condition $A^\mu A_\mu = -1$.  The \teves\ scalar is written as $\phi
= \phi_c + \delta \phi$. In the Newtonian limit at leading order we
take $\delta \phi << 1$, and again neglect time derivatives. One may
then verify that the disformal relation then relates these
perturbations: $\Phi = V+\phi$.

Consider first the scalar field equation at leading order in the
Newtonian expansion,
\begin{equation}
\label{scalarnonrel}
\grad\cdot\left[\mu \left(\kappa l^2 \left( \grad \delta\phi\right)^2\right) \grad \delta\phi\right]=\kappa G \tilde{\rho}.
\end{equation}
Note that we have not linearised the argument of the $\mu$ function in
$\delta \phi$ since while $\delta \phi << 1$, the \teves\ parameter
$l$ is precisely large enough to balance this - hence we may recover
MoND at leading order in the Newtonian analysis. There is no such
subtlety in the other field equations, and we may straightforwardly
perform a linearization in the Newtonian potentials.  All components
of the vector field equation vanish except for the $t$-component,
\begin{equation}
\lambda +\left(K-\frac{c_+}{2}\right) \grad^2 V = (-4\phi_c) 8\pi G\tilde{\rho},
\end{equation}
and similarly all components of Einstein's equation vanish with the
exception of the $tt$-component,
\begin{equation}
\lambda+(2+c_4-c_+)\grad^2 V = (1-8\phi_c) 8\pi G \tilde{\rho}.
\end{equation}
where we note that, as shown by Bekenstein \cite{Bekenstein:2004ne},
the term $\mathcal{F}$ doesn't contribute to the stress tensor in the
leading order Newtonian limit.  Combining the above results, we learn
how $\Phi$ is related to $\Phi_N$ and $\phi$,
\begin{equation}\label{ppprelation}
\Phi = \left(1+\frac{2K-2c_4+c_+}{4}-4\phi_c\right)\Phi_N +\phi.
\end{equation}
Hence, by following the same arguments that apply to pure TeVeS theory
(as discussed by Bekenstein in~\cite{Bekenstein:2004ne}), MoND
phenomenology results from (\ref{scalarnonrel}) and
(\ref{ppprelation}) when $c_4$ and $c_+$ are suitably
constrained. Hence our proposed modification of TeVeS, which likely
can avoid caustic formation for specific parameter ranges, does indeed
correctly reproduce the MoND limit which is the raison d'etre for
TeVeS.

\section{Summary and Discussion}\label{disc}

We have argued that Bekenstein's original formulation of TeVeS, while
reproducing MoND phenomenology, is actually dynamically badly
behaved. We have shown analytically and numerically in a variety of
situations that the integral curves of the vector field quickly and
generically evolve from regular initial data to caustic
singularities. Once this occurs the classical evolution to the future
is ill-posed. 

Since the time scale to a singularity is not suppressed
by any parameters, and is only determined by the initial data itself,
it seems the original formulation of TeVeS is unlikely to provide a realistic
theory of modified gravity. Put another way, while TeVeS does
reproduce Newtonian and Modified Newtonian dynamics in a non-relativistic regime, it appears that in many
cases this regime is unstable, the instability leading to the caustic
singularities. We stress that already the analytic arguments of
section \ref{analyticshock}, whilst made in the absence of matter,
already highlight instability in the dynamics of TeVeS. The latter
sections of the paper merely serve to illustrate that the situation
remains unchanged when one considers the dynamics in the presence of
matter. Indeed since matter can focus vector field curves towards it, it
can make the situation worse.

It is useful to contrast this situation with the singularities that
form in GR. We are very familiar with the fact that given certain
initial data, matter can collapse to a singularity in GR, on the time
scale of the gravitational in-fall time. However, in GR Cosmic
Censorship means that these singularities are always hidden behind
event horizons. Hence evolution outside the horizon is perfectly well
defined. In contrast, the vector field singularities we have exhibited
here lie outside any event horizon, and hence evolution is impossible
in the future null cone of these points unless there is some way to
understand these singularities beyond the TeVeS effective field
theory.

A similar situation arises in the perfect fluid, dust description of
dark matter. The evolution of such a fluid also forms caustics
(shocks) on time-scales of the in-fall time. Caustic formation in this
case signals the breakdown of the fluid description of the dark matter
and a requirement for a microscopic, particle description. Similarly
our results indicate that the dynamical regime where TeVeS can be
applied is limited by the breakdown of the effective theory on in-fall
timescales. In this case however we don't have a microscopic
description to transition to and the regime where the effective
description appears to break down is relevant to the motivation of the
theory itself.

Nonetheless, we believe a relatively minor modification, namely
generalizing the vector action of TeVeS to a form like that of
Einstein-\AE ther is likely to be able to give a dynamically well
defined theory, which as we have shown still gives MoND behaviour for
non-relativistic situations. It is interesting that a possibly related
instability was observed for linear perturbations about the
spherically symmetric static backgrounds of Giannios
\cite{Giannios:2005es} by Seifert \cite{Seifert:2007fr} and the same
modification was proposed, although the recovery of the MoND limit had
not previously been checked.

We emphasize that the detailed predictions of this modified TeVeS will
likely differ from the original TeVeS theory, and therefore any
phenomenological studies of TeVeS testing its ability to explain
astrophysical or cosmological data without dark matter should be
careful to include the necessary modification. It would be interesting
to revisit the questions addressed in
\cite{Skordis:2005eu,Koivisto:2008ig,Chen:2007ju,Xu:2007gk,Milgrom:2007br,Feix:2007zm,Schmidt:2007vj,Zhang:2007nk,Takahashi:2007nj,Famaey:2006iq,Bertolami:2006me,Bourliot:2006ig,Chen:2006vf,Chen:2006yd,Jin:2006sm,Bekenstein:2006fi,DiazRivera:2006qh,Zhao:2005xk,Chiu:2005ug,Skordis:2005xk,Dodelson06,Zhao:2005za,Slosar:2005fg}
using the modified theory.

We have given large classes of new black hole solutions in the
original TeVeS theory. However, these and the earlier solutions of
Giannios are not solutions of the modified TeVeS theory. Instead in
the case of modified TeVeS black hole solutions with a constant scalar
(and it is likely there are not `nearby' solutions with non-constant
scalar) will be identical to those in Einstein-\AE ther theory
discussed by Eling \& Jacobson \cite{Eling:2006ec,Eling:2006df}. In
particular there is no `charge' parameter $Q$, with the static black
hole geometries only being parametrized by one parameter, the mass
$M$. However the qualitative behaviour of the solutions is somewhat
similar to ours, with the integral curves of the vector field falling
into the future horizon. It would be very interesting to study the
geometry of the MF metric for these solutions. As a passing comment
Eling {\sl et al.} \cite{Eling:2007qd} have argued that such black holes may
violate the Generalised Second Law. It is interesting that while the
black hole solutions we found in the original TeVeS do not necessarily
violate the law since their entropy depends on multiple charges $M$
and $Q$, for black hole solutions in the modified TeVeS there is only
a one parameter family and the arguments of Eling {\sl et al.} apply.

\section*{Acknowledgements}

We wish to acknowledge Pedro Ferreira, Andrew Jaffe, and Jo\~ao
Magueijo for useful discussions. We also thank Jacob Bekenstein and Martin Feix for valuable comments.
TW is supported by an STFC advanced
fellowship and a Halliday award. BW is supported by an STFC
studentship. CRC acknowledges support from the Nuffield Foundation.

\appendix

\section{Approximately constant scalar solutions in the Newtonian regime}\label{argument}
In the main text we have made the claim that in the absence of matter we may formally consider solutions of TeVeS with constant scalar, and yet for the length scale of interest, let us call it $L$, which is much shorter than the TeVeS length scale $\ell$, we are still in the Newtonian regime, $\mu \simeq 1$. We envisage our scales of interest  $L$ to be of order planetary or solar system scales.
We claimed that to any solution with constant scalar, one can consider deforming the solution by adding scalar gradients that are tiny compared to our scale of interest $L$, and hence effectively negligible, but that would still be large enough over the region of interest to put the theory into the strong acceleration regime. The equation determining  $\sigma$ is,
\begin{equation}
- \mu {\cal F}(\mu) - \frac{1}{2} \mu^2 {\cal F}'(\mu) = y
\end{equation}
where we have written $\mu = \kappa G\sigma^2$ and $y = \kappa \ell^2 | \phi |^2$ with  $h^{\alpha\beta} = g^{\alpha\beta} - A^\alpha A^\beta$ and we have introduced the notation $| \phi |^2 \equiv h^{\alpha\beta} \phi_{,\alpha} \phi_{,\beta}$. Let us also choose, following Bekenstein \cite{Bekenstein:2004ne}, an ${\cal F}$ such that in the Newtonian limit $\mu \rightarrow 1$, we have,
\begin{equation}
{\cal F} \rightarrow \frac{3}{2} \ln(1-\mu) , \qquad y \rightarrow \frac{3}{4} \frac{1}{1 - \mu}
\end{equation}
and hence in this limit we have,
\begin{equation}
\mu = \kappa G \sigma^2 = 1 + O( \kappa \ell^2 | \phi |^2)^{-1} .
\end{equation}
We see a potential dilema in this claim is that naively $\mu \simeq 1, y \rightarrow \infty$ appears precisely at odds with a constant scalar which has $y = 0$. Hence how can a constant scalar solution ever be `close' to a solution in the strong acceleration regime $\mu \simeq 1$. The resolution is that what matters is not $| \phi |^2$, but $\ell^2 | \phi |^2$, and hence one can have a tiny gradient on scales of interest $L$ over a region of size $L$, but provided $\ell$ is large enough, one can still have $\mu \simeq 1$. We will now formalize this more carefully, by providing a prescription to take a solution of Einstein-Maxwell theory and then generating a solution of TeVeS with almost constant scalar in a controlled manner.

In the absence of matter the scalar equation reduces to,
\begin{eqnarray} 
\nabla_\beta \left( \mu(  \kappa \ell^2 | \phi |^2 ) h^{\alpha\beta} \phi_{,\alpha}
\right) &=& 0 ,
\end{eqnarray}
and the vector, which satisfyies the constraint $A^\mu A_\mu = -1$, obeys,
\begin{eqnarray} 
K \nabla_\beta F^{\beta \alpha} + \lambda A^{\alpha} +
\frac{8\pi}{\kappa} A^{\beta} \phi_{,\beta} g^{\alpha \gamma}
\phi_{,\gamma} = 0 ,
\end{eqnarray}
with the Einstein equations governing the metric becoming,
\begin{equation}
G_{\alpha \beta}=  \Theta^{scalar}_{\alpha \beta} + \Theta ^{vector}_{\alpha \beta} + \Theta ^{\mathcal{F}}_{\alpha \beta}  ,
\label{metrice2}
\end{equation}
with,
\begin{eqnarray}
\Theta ^{scalar}_{\alpha\beta} & \equiv & 8 \pi G \sigma^2\left[\phi_{,\alpha}\phi_{,\beta}-{ 1\over
 2}g^{\mu\nu}\phi_{,\mu}\phi_{,\nu}\,g_{\alpha\beta}- \right. \nonumber \\
&& \left. A^\mu\phi_{,\mu}\left(A_{(\alpha}\phi_{,\beta)}- { 1\over
 2}A^\nu\phi_{,\nu}\,g_{\alpha\beta}\right)\right]
\end{eqnarray}
\begin{equation} 
\Theta ^{vector}_{\alpha\beta}\equiv K\left(F_{\alpha}^{\p{\alpha}\mu}
F_{\beta \mu}-{ 1\over  4} g_{\alpha\beta} F^2
\right)- \lambda A_\alpha A_\beta
\end{equation}
\begin{equation}
\Theta ^{\mathcal{F}}_{\alpha\beta} \equiv - 2 \pi G^2 \ell^{-2}\sigma^4
{\cal F}(\kappa G\sigma^2)  g_{\alpha\beta}
\end{equation}
We consider the dimensionless constants $K, \kappa$ to be small but order one in what follows.

Let us now consider a solution, $\hat{A}_\alpha, \hat{g}_{\alpha\beta}$ of Einstein-Maxwell theory with gauge constraint $\hat{A}^\mu \hat{A}_\mu = -1$ imposed using the Lagrange multiplier $\lambda$ as for TeVeS,
\begin{eqnarray} 
&& \hat{G}_{\alpha\beta} = K\left( \hat{F}_{\alpha}^{\p{\alpha}\mu}
\hat{F}_{\beta \mu}-{ 1\over  4} \hat{g}_{\alpha\beta} \hat{F}^2 \right) - \lambda \hat{A}_\alpha \hat{A}_\beta \nonumber \\
&& K \hat{\nabla}_\beta \hat{F}^{\beta \alpha} + \hat{\lambda} \hat{A}^\alpha = 0 .
\end{eqnarray}
Here quantities with hats are composed from $\hat{A}_\alpha, \hat{g}_{\alpha\beta}$. Let us characterize the length scales of interest by the length $L$, and restrict our attention to a spacetime region $V$ of spatial size $\sim L$. Hence the curvatures of interest in the solution $(\hat{A}_\alpha, \hat{g}_{\alpha\beta})$ will scale as $1/L^2$ are we are not interested in much smaller curvature scales. We are envisaging that this scale is of order the solar system or less and compared to the TeVeS lengthscale $\ell$ we have a vast separation of scales,
\begin{eqnarray} 
L << \ell .
\end{eqnarray}
We begin by constructing a solution for a scalar $\hat{\phi}$ on the \emph{fixed} solution $(\hat{A}_\alpha, \hat{g}_{\alpha\beta})$ in the spacetime region $V$, of spatial scale $\sim L$. We take the scalar to obey the equation,
\begin{eqnarray} 
\hat{\nabla}_\beta \left( \hat{h}^{\alpha\beta} \hat{\phi}_{,\alpha}
\right) &=& 0 
\end{eqnarray}
where $\hat{h}^{\alpha\beta} = \hat{g}^{\alpha\beta} - \hat{A}^{\alpha}\hat{A}^{\beta}$, and we emphasize that we are ignoring any backreaction - it is simply a scalar on our fixed solution. We require that in our region $V$ the solution everywhere obeys the condition,
\begin{eqnarray} \label{condition}
1 >> L^2 \; \hat{h}^{\alpha\beta} \hat{\phi}_{,\alpha} \hat{\phi}_{,\beta} >> \frac{L}{\ell} .
\end{eqnarray}
For example, taking the trivial flat space solution $\hat{g}_{\alpha\beta} = \eta_{\alpha\beta}$, $\hat{A} = \frac{\partial}{\partial t}$, one might choose the scalar to be $\hat{\phi} = \alpha \, x$ where $x$ is one of the spatial coordinates, and $\alpha$ is a constant in the range $1 >> L^2 \alpha^2 >> L / \ell$. In general we expect to be able to find solutions obeying the condition \eqref{condition}, although we do not have a formal existance proof of this.

From this Einstein-Maxwell solution and the associated scalar solution $(\hat{\phi}, \hat{A}_\alpha, \hat{g}_{\alpha\beta})$ we may construct an approximate solution of the TeVeS equations, $({\phi},{A}_\alpha,{g}_{\alpha\beta})$, perturbatively in the dimensionless 
constant,
\begin{eqnarray} 
\epsilon & = & \left( \frac{L}{\ell} \right)^{1/4}
\end{eqnarray}
as,
\begin{eqnarray} \label{ansatz}
g_{\mu\nu} & = & \hat{g}_{\mu\nu} + \epsilon \, g^{(1)}_{\mu\nu}  \nonumber \\
A_{\mu} & = & \hat{A}_{\mu} + \epsilon \, A^{(1)}_{\mu} \nonumber \\
\lambda & = & \hat{\lambda} + \epsilon^2 \, \lambda^{(1)} \nonumber \\
\phi & = & \phi_0 + \epsilon \; \hat{\phi} + \epsilon^2 \, \phi^{(1)} ,
\end{eqnarray}
where $\phi_0$ is a constant and plays no role in the vacuum TeVeS equations which only involve $\phi$ derivatives. Taking the limit $\epsilon \rightarrow 0$, ie. looking at small scales compared to the TeVeS length scale $\ell$, we therefore see that the dynamics of TeVeS on scales $\sim L$ is given by precisely the Einstein-Maxwell solution $(\hat{A}_\alpha, \hat{g}_{\alpha\beta})$ and hence by an effectively (although not precisely) constant TeVeS scalar. We may think of the $\epsilon \rightarrow 0$ limit as fixing $L$ and taking $\ell$ to infinity, or alternatively and more physically for fixed $\ell$, focussing our interest on smaller and smaller length scales $L$.

Let us now check this claim. Firstly let us consider the scalar equation at leading order in the $\epsilon$ expansion. Consider the behaviour of $\mu( \kappa l^2 |\phi|^2 )$. From our condition \eqref{condition} above we see that,
\begin{eqnarray} 
1 >> \frac{ L^2}{\epsilon^2} \; \left( h^{\alpha\beta} {\phi}_{,\alpha} {\phi}_{,\beta} + O(\epsilon^2) \right) >> \frac{L}{\ell} .
\end{eqnarray}
so that for $\epsilon \rightarrow 0$ we have,
\begin{eqnarray} 
\ell^2 \; h^{\alpha\beta} {\phi}_{,\alpha} {\phi}_{,\beta} >>  \frac{1}{ \epsilon^{2} }
\end{eqnarray}
and thus we see that,
\begin{eqnarray} 
\mu( \kappa l^2 |\phi|^2 ) \simeq 1 - O(\epsilon^{2})
\end{eqnarray}
so that in our region $V$ we are forced into the Newtonian regime of TeVeS, even though the scalar field gradient is perturbatively small compared to the scale $L$ of interest in our region.
Hence the TeVeS scalar equation becomes,
\begin{eqnarray} 
\nabla_\beta \left( h^{\alpha\beta} \phi_{,\alpha} \right) &=& O(\epsilon^3)  ,
\end{eqnarray}
with the right-hand side coming from the non-constant part of $\mu$. This is indeed consistent with our ansatz \eqref{ansatz} above for the constant and $\hat{\phi}$ pieces with the correction term, $\phi^{(1)}$, accounting for the lower orders. $\phi^{(1)}$ is sourced primarily by the $O(\epsilon)$ corrections to $\nabla_\alpha$ and $h^{\alpha\beta}$, from the metric and vector corrections $A^{(1)}_\alpha$ and $g^{(1)}_{\alpha\beta}$, with the source from $\mu$ actually being sub leading to this.

We now show that just as the scalar equation is consistently solved perturbatively in $\epsilon$ by our ansatz, the Einstein and vector equations are too. In particular we must show that the backreaction in the Einstein equations from the scalar $\phi$ and TeVeS function $\mathcal{F}$ are small compared to the characteristic curvature scale $1/L^2$ in the solution $(\hat{g}_{\mu\nu}, \hat{A}_\mu)$.
Now following from our condition \eqref{condition} we have that the scalar in our region obeys the bound,
\begin{eqnarray} 
\frac{\epsilon^{2}}{L^2}  >> h^{\alpha\beta} {\phi}_{,\alpha} {\phi}_{,\beta} ,
\end{eqnarray}
and in addition we have an estimate for the contribution of $\mathcal{F}$ in the Einstein equations,
\begin{eqnarray} \label{ignoreF}
\Theta ^{\mathcal{F}}_{\alpha\beta} & \sim & \frac{1}{\ell^{2}}  \ln(1-\mu)  g_{\alpha\beta} \nonumber \\
& \sim & \frac{1}{L^{2}} \left( \epsilon^8 \ln \epsilon \right) \; \hat{g}_{\alpha\beta} .
\end{eqnarray}
Hence we see that the Einstein and vector equations to lowest order in $\epsilon$ reduce simply to the Einstein-Maxwell ones, and our ansatz \eqref{ansatz} therefore solves them to lowest order. The leading higher order corrections come from the perturbatively small backreaction of the scalar, and lead to the $O(\epsilon)$ corrections to $g_{\mu\nu}, A_\mu$, with the TeVeS function $\mathcal{F}$ essentially being negligible in the Newtonian limit as discussed in Bekenstein's original paper.

We have now more carefully justified our claim in the main text, namely that we may consider the TeVeS scalar to be effectively constant, and still be in the Newtonian regime $\mu \rightarrow 1$, provided we are restricting our interest to a region of scale $L << l$, as we are in the main text. Our first application is to consider a bound on caustic formation time using Raychaudhuri's equation for the Einstein-Maxwell system. Since caustic formation is local, we are only concerned with the spacetime in the region of scale $L$ where the caustic forms, and not the asymptotic behaviour of our geometry. The physical setting would be caustic formation on, for example, solar system scales $L$, with the gradient of the scalar arising from much larger galactic scales $\ell$. The second application is to embed the Einstein-Maxwell black holes in TeVeS. Again, since we are not interested in the far asymptotic region of these solutions, we may again employ our approximation to ignore scalar gradients and the TeVeS function $\mathcal{F}$ in the stress energy. For a region of size $L$ surrounding the black hole, the corrections will be characteristic scale $\ell$ and for any astrophysical black hole, phenomenologically this region of interest certainly obeys $L << \ell$.

We conclude with our previous example; $\hat{g}_{\alpha\beta} = \eta_{\alpha\beta}$, $\hat{A} = \frac{\partial}{\partial t}$, $\hat{\lambda} = 0$, and scalar solution on this background $\hat{\phi} = \alpha \, x$ with the constant $\alpha$ obeying $1 >> L^2 \alpha^2 >> L / \ell$. Hence we may take $\alpha = \epsilon/L$.
In this case the exact TeVeS solution can be found in the absence of the $\mathcal{F}$ term, which we have argued is subdominant  over the other corrections;
\begin{eqnarray} 
ds^2 & = & - a(x)^2 dt^2 + dx^2 + \frac{1}{a(x)} ( dy^2 + dz^2 ) \nonumber \\
A & = & \frac{1}{a(x)} \frac{\partial}{\partial t} \nonumber \\
\phi & = & \phi_0 + \epsilon \, \frac{x}{L}
\end{eqnarray}
with the function,
\begin{eqnarray} 
a(x) & = & A \, e^{\pm \epsilon {\frac{4\sqrt{\pi}}{  \sqrt{\kappa(2 K - 3)}}} \frac{x}{L}}
\end{eqnarray}
and one finds $\lambda = 16 K \pi \alpha^2/(\kappa (2 K - 3))$. Note that $y$ is actually a constant for this solution, $y = \kappa \, \epsilon^{-7/4}$ , so that $\mu$ is also constant.
Expanding in $\epsilon$ one finds this to be consistent with the ansatz. We have ignored the $\mathcal{F}$ term in this exact solution and a calculation confirms this is of order the estimate \eqref{ignoreF}, and hence vastly subdominant to the leading $\epsilon$ corrections in the solution.

\section{Constant scalar static black hole derivation}\label{constbhdetails}

We begin with a spherically symmetric, stationary system in
Schwarzschild coordinates for which,
\begin{eqnarray} 
\phi&=&\phi_c, \\
A^\mu&=&\left(A^t(r),A^r(r),0,0\right),\label{vectorcomps}\\
ds^2&=&-T(r)dt^2+R(r)dr^2+\nonumber\\
&&r^2\left(d\theta^2+\sin^2\theta d\varphi^2 \right),\label{metricansatz}
\end{eqnarray}
where $\phi_c$ is constant. The scalar field equation is
trivially satisfied for a vacuum spacetime. The vector field equation becomes
\begin{equation} K\nabla_\beta F^{\beta \alpha}+\lambda
  A^{\alpha}=0.\label{vec}.
\end{equation}
For the case where $\alpha=r$ in equation (\ref{vec}), the first term
vanishes, leaving the branch choice
\begin{equation}
\lambda A^r=0,
\end{equation}
thus for $A^r\not=0$ we have $\lambda=0$ everywhere. In this case the
field equations become those of Einstein-Maxwell theory for a
particular choice of gauge. Given this we expect to find
Reissner-Nordstr\"om black holes. Expressing the field equations in
terms of our metric and vector ans\"atze, we obtain three useful
components of the Einstein equations,
\begin{widetext}
\begin{eqnarray}
(tt)&
\frac{T}{Rr^2}\left(rR'+R^2-R\right)=\frac{K}{2}(A'_t)^2,\label{tt}\\
(rr)& \frac{rT'-RT+T}{r^2}=-\frac{K}{2}(A'_t)^2,\label{rr}\\ (\theta
\theta)&
\frac{1}{2RT}\left(2T'RT-2R'T^2-T'R'Tr+T''RTr-T'^2Rr\right)=Kr(A'_t)^2,\label{thth}
\end{eqnarray}
\end{widetext}
where primes indicate derivatives with respect to $r$.  Eliminating
$A'_t$ between equations (\ref{tt}) and (\ref{rr}) leads to the
relation
\begin{equation}
TR=C_1, \label{step1}
\end{equation}
where $C_1$ is a constant, which we set to 1 using the freedom
available in rescaling the $t$ coordinate at this stage. Performing
the same elimination between equations (\ref{rr}) and (\ref{thth}) and
substituting for $T$ using (\ref{step1}), we arrive at a solution for
the second metric component,
\begin{equation}
R=\left(1-\frac{2M}{r}+\frac{Q^2}{r^2}\right)^{-1},\label{R}
\end{equation}
with the corresponding solution for $T$, 
\begin{equation}
T=\left(1-\frac{2M}{r}+\frac{Q^2}{r^2}\right),\label{T} 
\end{equation} 
where $M$ and $Q^2$ are integration constants. The quantity $Q^2$ will
indeed turn out to be positive.

Consider the $t$ component of the co-vector
field equation 
\begin{eqnarray}
\frac{1}{2TR^2r}\left(2KrA''_tRT-KrA'_tT'R-KrA'_tR'T+\right.\nonumber\\
  \left.4KA'_tRT\right)=0.
\end{eqnarray} 
Note that the radial vector component appears nowhere in these
field equations, and will be determined algebraically using the field
equation for the Lagrange multiplier field, $\lambda$.  Substituting
in the metric components, and solving the resulting equation for $A_t$
we have
\begin{equation} 
A_t=C_2+\frac{C_3}{r},
\end{equation} 
where $C_2$ and $C_3$ are two more integration constants.

To determine the value of $C_2$ consider
the $\lambda$ equation, 
\begin{equation} \label{constraint}
\frac{A_r^2}{R}-\frac{A_t^2}{T}=-1,
\end{equation} 
as $r\rightarrow\infty$, $A_r$ must be driven to zero so that isotropy
is restored at spatial infinity. This expression therefore forces
$A_t^2\rightarrow 1$, and so we find $C_2=\pm 1$. We choose the
vector $A$ to be future pointing at spatial infinity, and so we pick
$C_2=-1$. The value of $C_3$ can then be determined straightforwardly
by substituting the expressions back into the Einstein equations. In
particular, for the $(\theta,\theta)$ component we find
\begin{equation} 
\frac{Q^2}{r^2}=\frac{KC_3^2}{2r^2}, 
\end{equation}
justifying our choice of the positive quantity $Q^2$. We then identify
\begin{equation} 
C_3=\sqrt{\frac{2}{K}}\, Q, 
\end{equation}
where $Q$ can be positive or negative. All that is left is to determine
$A_r$ through the constraint equation (\ref{constraint}) 
\begin{eqnarray}
A_r&=&\pm\left(1-\frac{2M}{r}+\frac{Q^2}{r^2}\right)^{-1}\nonumber\\
&&\sqrt{\left(\frac{2}{K}-1\right)\frac{Q^2}{r^2}+\left(M+\sqrt{\frac{2}{K}}Q\right)\frac{2}{r}}.
\end{eqnarray}
Phenomenologically $K< 1$ \cite{Bekenstein:2004ne}, so the first term
in the square root is positive. However at large $r$ the second term
is dominant and is possibly negative. Thus for $A_r$ to be real we
must satisfy the following bound
\begin{equation} 
M+\sqrt{\frac{2}{K}}Q\geq 0.
\end{equation} 

To summarise,
\begin{widetext}
\begin{eqnarray}
g&=&\mbox{Diag}\left[-\left(1-\frac{2M}{r}+\frac{Q^2}{r^2}\right),\left(1-\frac{2M}{r}+\frac{Q^2}{r^2}\right)^{-1},r^2,r^2
\sin^2(\theta)\right] \\ 
A_t&=&-1+ \sqrt{\frac{2}{K}} \frac{Q}{r}\\
A_r&=&\pm\left(1-\frac{2M}{r}+\frac{Q^2}{r^2}\right)^{-1}\sqrt{\left(\frac{2}{K}-1\right)\frac{Q^2}{r^2}+\left(M+\sqrt{\frac{2}{K}}Q\right)\frac{2}{r}},
\end{eqnarray} 
\end{widetext}
Thus the solution is Reissner-Nordstr\"om (RN), but where
the gauge freedom for the field $A$ has been fixed up to a sign by the
Lagrange multiplier constraint. 

We have obtained a black hole solution in the EF metric. Its horizons
are the ones observed by the gravitational components $g$, $A$, and
$\phi$. Any matter fields are influenced by the MF metric
(\ref{disformal}) and it is
important to consider the MF solution which is the observable
frame.

\section{Details of non-existence argument for black holes with
  non-constant scalar}\label{singularbhdetails}

In this appendix we will argue that there are no black hole solutions
`near' to those found in section \ref{constphisol} where the scalar is
not constant. To support this hypothesis we first consider small
perturbations to the scalar field about this constant scalar black
hole solution. We only need consider one equation
(\ref{scalarscalar}). Since the background value for the scalar field
is a constant, the equation of evolution for the scalar perturbation
is simply
\begin{equation} \delta\phi'=\frac{2 C_1 K}{2K(r-2)r + (K-1)\left(\frac{Q}{M}\right)^2-2K\sqrt{\frac{2}{K}}\frac{Q}{M} r}.
\end{equation} 
Unless $C_1=0$ this diverges for two values of $r$, which we denote
$r_{\rm sing1}$ and $r_{\rm sing2}$ where $r_{\rm sing2}\geq r_{\rm sing1}$. The
horizon positions are at $r_+$ and $r_-$ with $r_+\geq r_-$. It is
straightforward to show that if $K>0$ and $-1\leq \frac{Q}{M} \leq 1$
then $r_{\rm sing2} > r_+$ and so this singularity will occur outside the
outermost horizon. A scalar field singularity is usually a symptom of
a singular geometry.

Of course, the linear theory breaks down as the scalar perturbations
become large, and so while it indicates a singularity might form
outside the horizon if we try to deform the scalar from being
constant, it cannot be trusted. Hence we also solved numerically the
full non-linear theory with asymptotic data that is close to that of
the constant scalar RN solution, for particular parameters. The full
numerical solutions confirm that indeed a naked singularity is found,
as hinted at by the linear theory, with curvature invariants clearly
diverging at the singular point. We now describe this in detail.

Using the same metric ansatz (\ref{metricansatz}) and the general form
for the vector field (\ref{vectorcomps}), the $r$-component of the
vector field equation is 
\begin{equation}
A_r\left(\lambda+\frac{8\pi(\phi')^2}{\kappa R}\right)=0,
\end{equation} 
and so for the case $A_r\not=0$, this equation determines $\lambda$ to
be proportional to the square of the proper derivative of
$\phi$. Substituting $A_r$ from the constraint equation
(\ref{constraint}), and this value of $\lambda$ into the other field
equations yields
\\
Metric ($t$,$t$)-component
\begin{eqnarray}
  -\frac{2T}{r}+\frac{2RT}{r}+\frac{2TR'}{R}-\frac{16\pi r T
    (\phi')^2}{\kappa}=&&\nonumber\\
  Kr(A'_t)^2+\frac{8\pi r A_t^2
    (\phi')^2}{\kappa}.&&\label{scalartt} 
\end{eqnarray} 
Metric ($r$,$r$)-component
\begin{equation} \frac{2}{r}-\frac{2R}{r}+\frac{2T'}{T} + \frac{K r
    (A'_t)^2}{T}-\frac{16\pi (\phi')^2}{\kappa}+\frac{8\pi r A_t^2
    (\phi')^2}{\kappa T}=0.\label{scalarrr} 
\end{equation} 
Metric ($\theta$,$\theta$)-component 
\begin{eqnarray} 
\frac{2r T
    R'}{R}-2rT'+\frac{r^2 R' T'}{R}+\frac{r^2 (T')^2}{T}+2K r^2
  (A'_t)^2&&\nonumber\\
-\frac{32\pi r^2 T (\phi')^2}{\kappa} + \frac{16\pi r^2
    A_t^2 (\phi')^2}{r}-2 r^2 T''=0.&&\label{scalarthth} 
\end{eqnarray}
Vector $t$-component 
\begin{equation}
  K\left(-\frac{4}{r}+\frac{R'}{R}+\frac{T'}{T}\right)A'_t+\frac{16\pi
    A_t (\phi')^2}{\kappa}=2KA''_t.\label{scalart} 
\end{equation}
Scalar 
\begin{equation}
  \phi'=\frac{C_1\sqrt{RT}}{r^2(2T-A_t^2)},\label{scalarscalar} 
\end{equation}
where $C_1$ is an integration constant from the second order scalar
field equation. This is normally associated with a scalar
mass. Eliminating $A_t$ and $A'_t$ between (\ref{scalartt}) and
(\ref{scalarrr})
\begin{equation}\label{combo1} 
\frac{16\pi r (\phi')^2}{\kappa}=\frac{(TR)'}{TR}, 
\end{equation} 
and eliminating the same variables between (\ref{scalarthth}) and
(\ref{scalartt}), then substituting for $\phi'$ using (\ref{combo1})
gives
\begin{equation}
  2T\left(\frac{2(R-1)}{r}+\frac{R'}{R}\right)+\left(\frac{rR'}{R}-6\right)T'+\frac{r(T')^2}{T}=2rT'',
\end{equation} 
which can be integrated once to obtain a first order equation in
$T$ 
\begin{equation} 
RT(C_2+4r^2 T)=4r^2T^2 + 4r^3 T T'  +r^4(T')^2, 
\end{equation} 
where $C_2$ is a constant of integration. This equation allows
elimination of one of the metric components $R$ from the system of
equations to numerically integrate. Choosing to eliminate $R$ and
$\phi'$ using (\ref{scalarscalar}), from (\ref{combo1}) and
(\ref{scalarrr}), gives the system of equations
\begin{widetext}
  \begin{eqnarray}
    T''&=&-\frac{2}{r(C_2+4r^2T)}\left(\frac{C_2T}{r}+2(C_2+2r^2T)T'-r^3T'^2-\frac{4C_1^2\pi(2T+rT')^3}{r\kappa(-2T+A_t^2)^2}\right),\label{tdiff}\\
    A'_t&=&\pm\sqrt{2}\left(\frac{(16\pi C_1^2-\kappa C_2(2T-A_t^2))(T+rT')+r^2(4\pi C_1^2 + r^2\kappa (2T-A_t^2))(T')^2}{r^2\kappa K (C_2 + 4r^2T)(2T-A_t^2)}\right)^{\frac{1}{2}}.\label{udiff}
  \end{eqnarray}
\end{widetext}
From which $\phi'$ can be obtained during the numerical integration
through the scalar field equation (\ref{scalarscalar}).

We then integrate inwards from large $r$, imposing an asymptotically
flat spacetime, a constant scalar and vanishing $A_r$ component at
spatial infinity. Looking first to the asymptotic expansion of $T$ and
$A_t$ will allow identification of the free parameters available for
the boundary condition at large $r$. Assuming the general asymptotic
form
\begin{eqnarray}
T(r)&=&t_0+\frac{t_1}{r}+\frac{t_2}{r^2}+ \mathcal{O}\left({\frac{1}{r^3}}\right),\\
A_t(r)&=&u_0+\frac{u_1}{r}+\frac{u_2}{r^2}+ \mathcal{O}\left({\frac{1}{r^3}}\right).\label{Atnumerics}
\end{eqnarray}
We may use the freedom in rescaling the $t$ coordinate in setting
$t_0=1$, and the freedom in the $r$ coordinate to set $t_1$ to
$-1$. Note that this explicitly sets the standard RN mass to be
positive. It then follows from the constraint equation that if we are
to have $A_r\rightarrow 0$ as $r\rightarrow\infty$ then
$A_t\rightarrow -1$ and $u_0=-1$. The remaining coefficients may be
determined by performing a series expansion about infinity of the two
differential equations (\ref{tdiff},\ref{udiff}). We find
\begin{eqnarray}
T& \simeq & 1-\frac{1}{r}+\frac{1+\frac{32\pi C_1^2}{\kappa} -C_2}{4r^2},\\
A_t &\simeq &-1+\pm\sqrt{\frac{2}{K}} \frac{\sqrt{1+\frac{16\pi
      C_1^2}{\kappa} -C_2}}{2r}- \frac{4\pi C_1^2}{\kappa
  K}\frac{1}{r^2},\nonumber\\
\,
\end{eqnarray}
where the sign choice comes from the sign of the gradient of $A_t$,
equation (\ref{udiff}). There is only one consistent choice for this,
as may be seen by using the Lagrange multiplier equation
(\ref{constraint}) to calculate the corresponding asymptotic expansion
for $A_r^2$. To lowest order:
\begin{equation}
A_r^2 = \left[1-\pm\sqrt{\frac{2}{K}} \sqrt{1+\frac{16\pi C_1^2}{\kappa} -C_2}\right]\frac{1}{r} + \mathcal{O}\left(\frac{1}{r^2}\right).
\end{equation}
Thus for the reality of $A_r$ we are forced to choose the negative sign. 

We are now left with four parameters: two constants, $\kappa$ and $K$
and two free parameters $C_1$ and $C_2$ with an additional condition
to ensure the reality of $A_t$
\begin{equation}
1+\frac{16 \pi}{\kappa} C_1^2-C_2>0.\label{realitycond}
\end{equation}
Additionally, one can identify these parameters with the standard
Reissner-Nordstr\"om charge to mass ratio,
\begin{equation}
\left(\frac{Q}{M}\right)^2 \equiv 1 -C_2,
\end{equation}
and a quantity associated with the scalar
\begin{equation}
S^2\equiv\frac{16\pi C_1^2}{\kappa},
\end{equation}
so that the reality condition (\ref{realitycond}) is now
guaranteed. Using this definition of $S$, the differential equations
to numerically integrate (\ref{tdiff},\ref{udiff}) are reduced to a
constant $K$, and the two parameters $S$ and $Q/M$.

Naively considering the $\phi$ equation of motion
(\ref{scalarscalar}), one sees that a possible divergence occurs when
$2T=A_t^2$. This does occur, and we demonstrate below for the
parameter choices $K=0.01,\,\kappa=0.01,\,Q/M=0.01$ and a small scalar
charge, $S=0.001$.

\begin{figure}[t]
\begin{center}
\includegraphics[angle=0,width=3.in]{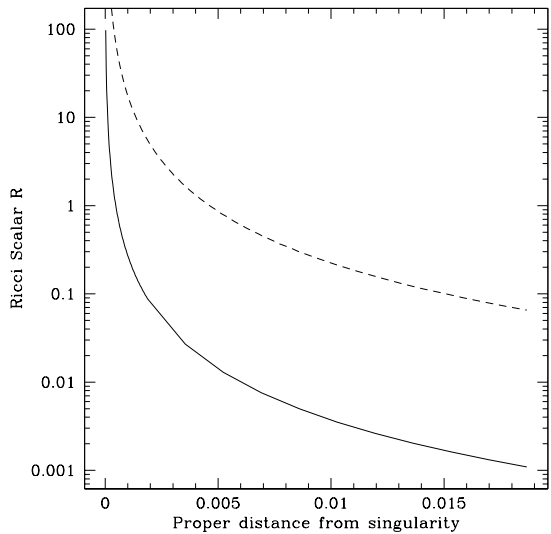}
\caption{EF (solid) and MF (dashed) Ricci Scalars
  for the parameters $S = 0.001$ and $K=\kappa=Q/M=0.01$ as functions
  of the proper distance from the singular surface, illustrating a
  physical singularity.}
\label{divergentphi1}
\end{center}
\end{figure}

\begin{figure}[t]
\begin{center}
\includegraphics[angle=0,width=3.in]{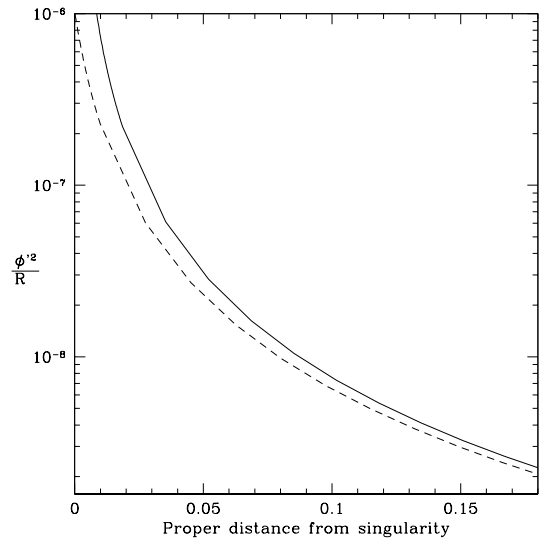}
\caption{Square of the proper derivative of $\phi$,
  $\frac{\phi'^2}{R}$, for the full integration (solid) and for the
  linearised theory (dashed). A larger $S$ would see a larger
  discrepancy between the singular positions for the full and the
  linearised theory. This plot also demonstrates the agreement of the
  two curves away from the singularity.}
\label{divergentphi2}
\end{center}
\end{figure}

We find that both the EF and MF Ricci scalars diverge, and thus it is
not a coordinate singularity - this is shown in the left panel of
figures~\ref{divergentphi1} and~\ref{divergentphi2}. Further, at no point outside of the
singular position does the $g_{tt}$ component of the metric vanish -
the singularity is not enclosed within a horizon. $\phi$, $T$ and
$A_t$ remain finite (though their proper gradients diverge) up to the
singular point. Interestingly, at the singular point $g_{rr}\equiv
R=0$ implying (through $2T-A_t^2=0$ and $A^2=-1$) that the radial
component of the TeVeS vector vanishes at this point.

We have shown that linear theory suggests no deformation of the
constant scalar RN solution that has regular horizon, and non-trivial
scalar. In specific cases we have confirmed that the full non-linear
theory agrees with the linear theory in that attempts to make the
scalar be non-constant lead to a naked singularity rather than a
regular horizon. However, we have not shown that there is no such
solution far in `solution-space' from the constant scalar RN
solution. We think it unlikely, although have not explored this
possibility in detail.

\section{Details of full dynamical numerical simulation method}\label{numericalmethod}

We use the following Schwarzschild-like coordinate system for our evolution,
\begin{equation}
ds^2=-T^2(t,r)dt^2+e^{R(t,r)}dr^2+r^2d\Omega_2^2 .
\end{equation}
In this coordinate system the Einstein equations have the usual
constraints (the $tt$ and $tr$ components) together with a second
order evolution equation for $R$, the $\theta \theta$ component. We do
not use this directly to evolve $R$, instead we use the $tr$
constraint equation itself. Once all variables (apart from $T$) have
been successfully evolved to the next spatial slice, we integrate
across the grid in the radial direction to obtain $T$ on the slice
using the $rr$ equation. A simplification is made for this radial
integration of $T$; we use the value of $T$ from the old spatial slice
to compute the contribution to the right hand side of the
$rr$-Einstein equation, for the sake of computational run time. Using
this approximation, the $rr$ component is a first order differential
equation in $T$.

We find that the most stable way to evolve the vector field is to
\emph{evolve} $\lambda$, rather than evolving $A$ and then calculating
$\lambda$ through contraction of the vector field equation. The
evolution equation we use for $\lambda$ is given by taking the
divergence of the vector equation
\begin{widetext}
\begin{equation}
  \nabla_\alpha \nabla_\beta F^{\beta \alpha} + \nabla_\alpha (\lambda A^\alpha) + \nabla_\alpha(\sigma^2 A^\beta \phi,_\beta g^{\alpha \gamma} \phi,_\gamma) = \nabla_\alpha \left[ (1-e^{-4\phi}) g^{\alpha \mu} A^\beta \Tilde{T}_{\mu \beta} \right].
\end{equation}
\end{widetext}
The first term vanishes through antisymmetry of $F$ and the symmetry
of the Ricci Tensor. The third term will contain second time
derivatives of $\phi$, and so we substitute for this using the $\phi$
equation of motion. The last term will contain second derivatives of
the $\chi$ field. For this we do not substitute from the $\chi$ field
equation as we are able to use the conservation properties of
$\tilde{T}$ to obtain a simpler (and so numerically advantageous)
expression. Consider the definition of $\tilde{T}$ from a variation of
the action with respect to the MF metric,
\begin{equation}\label{ttildedef}
\delta S = -\frac{1}{2} \tilde{T}_{\alpha \beta} \sqrt{\tilde{g}}\delta \tilde{g}^{\alpha \beta},
\end{equation}
and one may rewrite the variation of $\tilde{g}$ in terms of
variations of the other fields, through the disformal relation
\ref{disformal} (see \cite{Bekenstein:2004ne})
\begin{widetext}
\begin{equation}
\delta \tilde{g}^{\alpha \beta} = e^{2\phi} \delta g^{\alpha \beta} + 2 \sinh 2\phi A_\mu \delta g^{\mu (\alpha} A^{\beta )} + 2 \left[ e^{2\phi} g^{\alpha \beta} + 2 A^\alpha A^\beta \cosh 2 \phi \right] \delta \phi + 2 \sinh 2 \phi A^{(\alpha} g^{\beta ) \mu} \delta A_\mu.
\end{equation}
\end{widetext}
Specifically, consider the case of a diffeomorphism generated by the
vector field itself,
\begin{eqnarray}
\delta g^{\alpha \beta}  =  \pounds_A g^{\alpha \beta} &=& \nabla^{(\alpha} A^{\beta)},\\
\delta A^\mu = \pounds_A A^\mu &=& 0,\\
\delta \phi = \pounds_A \phi &=& A^\alpha \nabla_\alpha \phi,
\end{eqnarray}
so that
\begin{equation}
\delta A_\nu = \delta ( g_{\nu\mu} A^\mu) = - A_\alpha g_{\nu\beta} \delta g^{\alpha\beta} = - A_\alpha g_{\nu\beta} \nabla^{(\alpha} A^{\beta)}.
\end{equation}

Inserting all of this into (\ref{ttildedef}) and integrating by parts
from the terms containing variations of the inverse EF metric, we
obtain the following expression
\begin{equation}
A^\beta \nabla^\alpha \tilde{T}_{\alpha \beta} = \left( g^{\alpha \beta} + \left(1+e^{-4 \phi}\right) A^\alpha A^\beta \right) A^\mu \nabla_\mu \phi,
\end{equation}
which we can then use instead of the $\chi$ evolution equation in the
evolution of $\lambda$.

Thus both the metric components and $\lambda$ are evolved with first
order differential equations, whilst $A$, $\phi$ and $\chi$ are
evolved at second order. The origin boundary conditions are
$\phi,_r=0$, $A^r = 0$, $T=1$, $R=0$,$T,_r=0$, $R,_r=0$ and
$\chi,_r=0$. 

For the scalar shell collapse in section (\ref{bh2}) the initial conditions on the $t=0$ spatial slice, are
$\phi=\phi_c$, $A^r=0$, $R,_t=T,_t=\phi,_t=\chi,_t=0$. We consider two simulations with differing initial conditions for $F_{\mu\nu}$, and these are discussed in the text. The initial $R$ and $T$ configurations
are specified by solving the $rr$ and $tt$ Einstein equations for a
choice of Gaussian initial data on $\chi$. 
The $tr$-Einstein equation
is automatically satisfied for this initial data.

For the boson star in section (\ref{star}) the initial conditions for all fields are fixed by requiring for a static star configuration initially. This configuration was found using a radial shooting method, as discussed in the text.

Second order finite
differencing was used to discretize the equations of motion. For the
scalar-shell collapse simulation, we used resolutions up to $\Delta
r=0.05$ and $\Delta t=0.0005$, whilst for the Boson star simulation we
used resolutions up to $\Delta r=0.05$ and $\Delta t=0.00005$. Each
simulation had $400$ simulation sites per slice, and took on the order
of a day to run using an average desktop computer.

\section{Constraint Testing}\label{constrainttesting}

\begin{figure}[th]
\begin{center}
\includegraphics[width=3in]{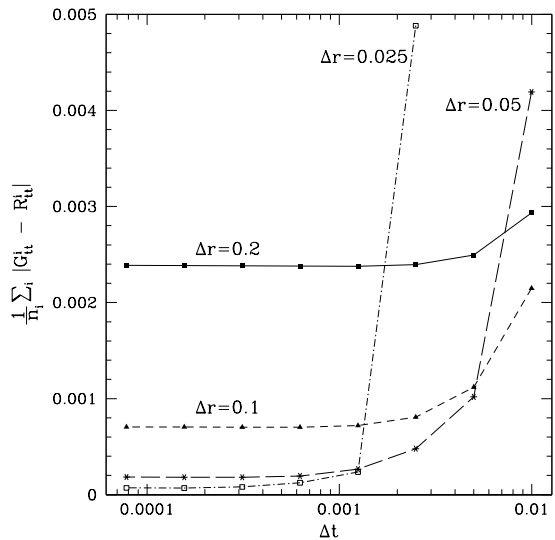}
\caption{The average of the $tt$ constraint equation over simulated
  grid points for the black hole of section \ref{bh2}, for different
  time steps. Each curve corresponds to a different spatial
  resolution.}
\label{ttcheck}
\end{center}
\end{figure}

We ran simulations at a variety of resolutions to check convergence
which was seen in accord with our second order finite differencing. To
test the constraint equations (and convergence), we consider the black
hole dynamics for some representative initial data as in section
\ref{bh2}, and take the absolute value of the constraint equation
(l.h.s. - r.h.s.) at a grid point, and then average this over all grid
points (labelled by $i$) in a large physical area. We then compute
this average for various spatial and temporal resolutions, keeping the
physical region fixed. We denote the set of all grid points in this
region as $\sigma$. Figure \ref{ttcheck} shows this sum for the $tt$
component of the Einstein equation against $\log_{10}\Delta t$, for
four spatial steps $\Delta r = 0.2$, $0.1$, $0.05$ and $0.025$. We
clearly see agreement with expected constraint behaviour for second
order differencing as the spatial and temporal resolutions are
reduced. Similar checks were performed for the boson star simulation
of section \ref{star}.

\end{document}